\documentclass[12pt,letterpaper]{article}
\pdfoutput=1
\usepackage{graphicx,array}
\usepackage{color}
\usepackage{latexsym}
\usepackage{amsthm}
\usepackage{amsmath}
\usepackage{amssymb}

\usepackage{dsfont}
\usepackage{epsfig}
\usepackage{slashed}
\usepackage{bbold}
\usepackage{psfrag}
\usepackage[svgnames]{xcolor}
\PassOptionsToPackage{caption=false}{subfig}
\usepackage{subcaption}
\usepackage{xfrac}
\usepackage{multirow}
\usepackage{booktabs}
\usepackage[colorlinks=true,linkcolor=MediumBlue,citecolor=Green,urlcolor=violet]{hyperref}
\usepackage{cite}

\newcommand{\be}{\begin{equation}}
\newcommand{\ee}{\end{equation}}
\newcommand{\bea}{\begin{eqnarray}}
\newcommand{\eea}{\end{eqnarray}}

\def\({\left(}
\def\){\right)}

\usepackage{soul}

\begin{document}

\begin{center}
{\LARGE \bf Phases of Cannibal Dark Matter}
\bigskip\vspace{1cm}{

Marco Farina$^1$, Duccio Pappadopulo$^2$, Joshua T. Ruderman$^2$, and Gabriele Trevisan$^2$}
\\[7mm]
 {\it 
$^1$New High Energy Theory Center, Department of Physics, Rutgers University,  \\
136 Frelinghuisen Road, Piscataway, NJ 08854, USA \\
$^2$Center for Cosmology and Particle Physics, Department of Physics, \\ 
New York University, New York, NY 10003, USA
 }

\end{center}

\bigskip \bigskip \bigskip \bigskip

%%%%%%%%%%%%%%%%%%%%%%%%%%%%%%%%%%%%%%%%%%%%%%%%%%%%%%%%%%%%%%%%%%%%%%%%%%
\centerline{\large\bf Abstract} \bigskip
A hidden sector with a mass gap undergoes an epoch of cannibalism if number changing interactions are active when the temperature drops below the mass of the lightest hidden particle.  During cannibalism, the hidden sector temperature decreases only logarithmically with the scale factor.  We consider the possibility that dark matter resides in a hidden sector that underwent cannibalism, and has relic density set by the freeze-out of two-to-two annihilations.  We identify three novel phases, depending on the behavior of the hidden sector when dark matter freezes out.  During the {\it cannibal} phase, dark matter annihilations decouple while the hidden sector is cannibalizing.  During the {\it chemical} phase, only two-to-two interactions are active and the total number of hidden particles is conserved.  During the {\it one way} phase, the dark matter annihilation products decay out of equilibrium, suppressing the production of dark matter from inverse annihilations.  We map out the distinct phenomenology of each phase, which includes a boosted dark matter annihilation rate, new relativistic degrees of freedom, warm dark matter, and observable distortions to the spectrum of the cosmic microwave background.
\bigskip\\
%%%%%%%%%%%%%%%%%%%%%%%%%%%%%%%%%%%%%%%%%%%%%%%%%%%%%%%%%%%%%%%%%%%%%%%%%%

\newpage

%%%%%%%%%%%%%%%%%%%%%%%%%%%%%%%%%%%%%%%%%%%%%%%%%%%%%%%%%%%%%%%%%%%%%%%%%%%%%%%%%%%%%%%%%%%%%%%%%%%%%%%%%%%%%%%%%%%%%%%%%%%%%%%%%%%%%%%%%%%%%%%%%%%%%%%%%%%%%%%%%%%%%%%%%%%%%%%%%%%%%%%%%%%%%%%%%%%%%%%%%%%%%%%%%%%%%%%%%%%%%%%%%%%%%%%%%%%%%%%%%%%%%%%%%%%%%%%%%%%%%%%%%%%%%%%%%%%%%%%%%%%%%%%%%%%%%%%%%%%%%%%%%%%%%%%%%%%%%%%%%%%%%%%%%%%%%%%%%%%%%%%%%%%%%%%%%%%%%%%%%%%%%%%%%%%%%%%%%%%%%%%%%%%%%%%%%%%%%%%%%%%%%%%%%%%%%%%%%%%%%%%%%%%%%%%%%%%%%%%%%%%%%%%%%%%%%%%%%%%%%%%%%%%%%%%%%%%%%%%%%%%%%%%%%%%%%%%%%%%%%%%%%%%%%%%%%%%%%%
\section{Introduction and Conclusions}
\label{sec:intro}
%%%%%%%%%%%%%%%%%%%%%%%%%%%%%%%%%%%%%%%%%%%%%%%%%%%%%%%%%%%%%%%%%%%%%%%%%%
According to the Weakly Interacting Massive Particle (WIMP) paradigm, the Dark Matter (DM) component of our Universe is composed of non-relativistic particles whose abundance is set by their annihilations in the early Universe~\cite{Lee:1977ua,Vysotsky:1977pe,Kolb:1990vq}. At high temperatures ($T>m_{DM}$), DM particles are assumed to be in thermal equilibrium with the Standard Model (SM) plasma. As the temperature drops below their mass, their abundance starts to decrease exponentially and number changing annihilation processes like ${\textrm{DM}}\,{\textrm{DM}}\leftrightarrow {\textrm{SM}}\, {\textrm{SM}}$ become inefficient. Eventually, when $T/m_{DM}\lesssim 1/{30}$, the DM comoving number density \emph{freezes out}. The resulting energy density is determined by the DM annihilation cross section, $\left< \sigma v \right>$,
\be \label{eq:VanillaOmega}
\Omega_{DM}h^2\simeq 0.1\frac{ (20\,{\textrm{TeV}})^{-2}}{\langle\sigma v\rangle}.
\ee 
This result implies that a particle with weak scale mass and electroweak size interactions ($\left< \sigma v \right>  \approx (20~\mathrm{TeV})^{-2} \approx 3 \times 10^{-26}~\mathrm{cm}^3/\mathrm{s}$) has a freeze-out abundance that matches the observed relic density: $\Omega_{DM}h^2\approx 0.1$~\cite{Ade:2015xua}.

Such a framework has various appealing features. The freeze-out mechanism is insensitive to any initial condition or UV physics due to its thermal nature. If DM mainly annihilates into SM particles, then it must have sizable interactions with the SM, opening up exciting experimental possibilities for its observation. Furthermore, the existence of new weak scale particles is motivated by theoretical considerations such as solving the naturalness problem of the Higgs mass.

However, recent experimental results are challenging this picture: DM direct detection experiments have excluded significant WIMP parameter space~\cite{Cushman:2013zza,Akerib:2015rjg,Aprile:2015uzo}, and are getting close to the neutrino background~\cite{Billard:2013qya}.  At the same time, collider and precision searches are increasingly constraining the possible existence of new physics around the weak scale.  Pending a discovery of a WIMP, it is highly motivated to explore new theories of dark matter and their experimental tests.

One possibility is that DM is part of a hidden sector of particles that is very weakly coupled to the SM\@ (see for example Refs.~\cite{Kolb:1985bf,Goldberg:1986nk,Carlson:1992fn,Strassler:2006im,Finkbeiner:2007kk,Pospelov:2007mp,Feng:2008ya,Feng:2008mu,ArkaniHamed:2008qn,Kaplan:2009ag,Cheung:2010gj,Hochberg:2014dra,D'Agnolo:2015koa,cannibal}). At sufficiently high temperatures, interactions within the dark sector  will guarantee thermal equilibrium among the dark sector particles. However, if interactions between the dark sector and the SM are sufficiently weak, the dark sector temperature $T_D$ and entropy $s_D$ will in general be different from the SM ones~\cite{Carlson:1992fn,Feng:2008mu,Cheung:2010gj,cannibal}. The fact that the DM is in equilibrium with a thermal bath while its energy density evolves to the observed value implies the same  insensitivity to UV dynamics as for standard WIMPs (except for sensitivity to the initial condition that sets the relative temperatures of the SM and dark sector plasmas).

\begin{figure}[!!!t]
\begin{center}
\includegraphics[width=0.4\textwidth]{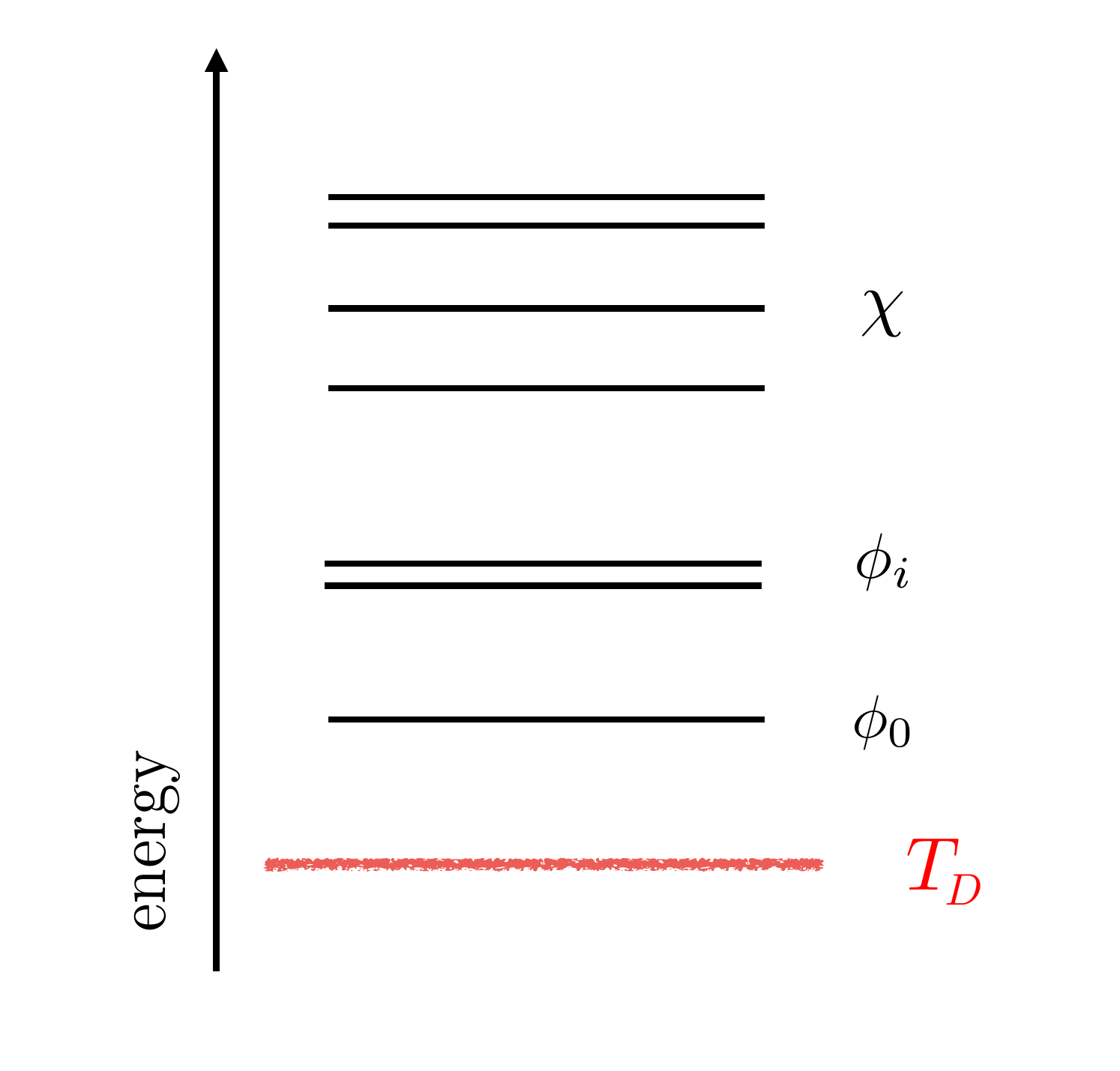}
\end{center}
\vspace{-.3cm}
\caption{\small { The typical mass spectrum of a gapped hidden sector. $\chi$ plays the role of the DM, which we assume to be stabilized by some symmetry. The various $\phi_i$ are generically unstable. $\phi_{0}$ is the Lightest Dark sector Particle (LDP) and has nonzero mass.  The hidden sector will undergo an epoch of cannibalism if the temperature drops below the mass of the LDP, $T_d < m_0$, while number changing interactions are still in equilibrium.}}
\label{fig:spectrum}
\end{figure}

Most works studying hidden sector dark matter assume that  the hidden sector contains relativistic particles in thermal equilibrium with DM when its annihilations freeze-out.  An alternative possibility is that DM resides in a hidden sector with a mass gap set by the mass of the Lightest Dark sector Particle (LDP), $m_0$.  If the hidden sector is sufficiently weakly coupled to the SM,  this opens up the possibility that DM is not in thermal contact with radiation when its annihilations decouple.  This implies that the hidden sector undergoes an epoch of cannibalism~\cite{Carlson:1992fn}, during which its temperature decreases only logarithmically with the scale factor. This possibility was first studied by Ref.~\cite{Carlson:1992fn}, which considers the possibility that DM freezes out through 3-to-2 annihilations.  Recently, some of us proposed that dark matter may reside in a hidden sector with a mass gap and have abundance that follows from 2-to-2 annihilations~\cite{cannibal}.  For additional studies that include cannibalism see Refs.~\cite{deLaix:1995vi,Boddy:2014qxa,Yamanaka:2014pva,Garcia:2015loa,Bernal:2015ova,Bernal:2015xba,Kuflik:2015isi,Soni:2016gzf,Forestell:2016qhc}.  

In this paper, we consider the following framework~\cite{cannibal}:
\begin{itemize}
\item {\bf Cannibal Dark Matter:} the abundance of DM is set by the freeze-out of 2-to-2 annihilations in a hidden sector  that undergoes an epoch of cannibalism that begins before DM annihilations decouple.
\end{itemize}
Our goal is to map out the different possible cosmologies for Cannibal DM, in order to identify viable models that reproduce the observed relic density.  We will find that gapped hidden sectors have a rich phase structure, with multiple novel avenues for DM freeze-out.

We assume that dark matter resides in a hidden sector with a mass gap, and that the hidden sector is kinetically decoupled from the SM such that it evolves with an independent temperature.  The hidden sector contains a stable particle, $\chi$, that will constitute DM, as well as massive unstable particles $\phi_i$ (see Fig.~\ref{fig:spectrum}). We can identify three relevant timescales in the cosmological evolution of the hidden sector:
\begin{itemize}

\item $t_f$ (and the corresponding dark temperature $T_f$)--- the time after which the rate of DM number changing processes, $\Gamma_f$, is smaller than the Hubble constant, $H$,
\be
\Gamma_f \equiv n_\chi^{eq} \left< \sigma_2 v \right> < H,
\ee
where $\sigma_2$ is the cross section for $\chi \chi \rightarrow \phi \phi$ (which is the leading process that changes DM number density) and $n_\chi^{eq}$ is the equilibrium number density of $\chi$. Here and below the indices on the $\phi_i$ are implied.  In general, $n_\chi^{eq} \propto e^{-(m_\chi-\mu_\chi)/T_D}$ includes a chemical potential, $\mu_\chi$, which as we will see below can play an important role.  After $t_f$, the comoving DM density is conserved. 

\item $t_c$ (and the corresponding dark temperature $T_c$)---  the time after which the rate of number changing processes in the hidden sector, $\Gamma_c$, falls below $H$.  This timescale is set by the decoupling of the final $n \leftrightarrow m$ process with $n \ne m$.  For example, $t_c$ can be set by the decoupling of  $\phi \phi \phi \rightarrow \phi \phi$ with cross section $\sigma_3$,
\be
\Gamma_c \equiv (n_\phi^{eq})^2 \left< \sigma_3 v^2 \right> < H.
\ee
After the time $t_c$, the total comoving number of particles in the hidden sector is conserved.  However, 2-to-2 processes may still be active and can change relative particle abundances.
 A chemical potential is necessary to enforce the conservation of the total comoving number of hidden sector particles. 

%%%%
\item $t_d$ (and the corresponding dark temperature $T_d$)---  the time at which the DM annihilation products, $\phi$, decay more rapidly than the Hubble time ({\it i.e.} the Universe becomes older than the $\phi$ lifetime), 
\be
\Gamma_\phi > H.
\ee
  As we discuss below, metastability of $\phi_i$ is a necessary condition for $\chi$ to constitute the totality of the observed DM abundance~\cite{cannibal}. If this was not the case, the lightest of the $\phi_i$, being non-relativistic, would dominate the dark sector energy density.
\end{itemize}
%%%%%%%%%%%%%
%%%%%%%%%%%%%
%%%%%%%%%%%%%
%%%%%%%%%%%%%
\begin{figure}[!!!t]
\begin{center}
\includegraphics[width=0.75\textwidth]{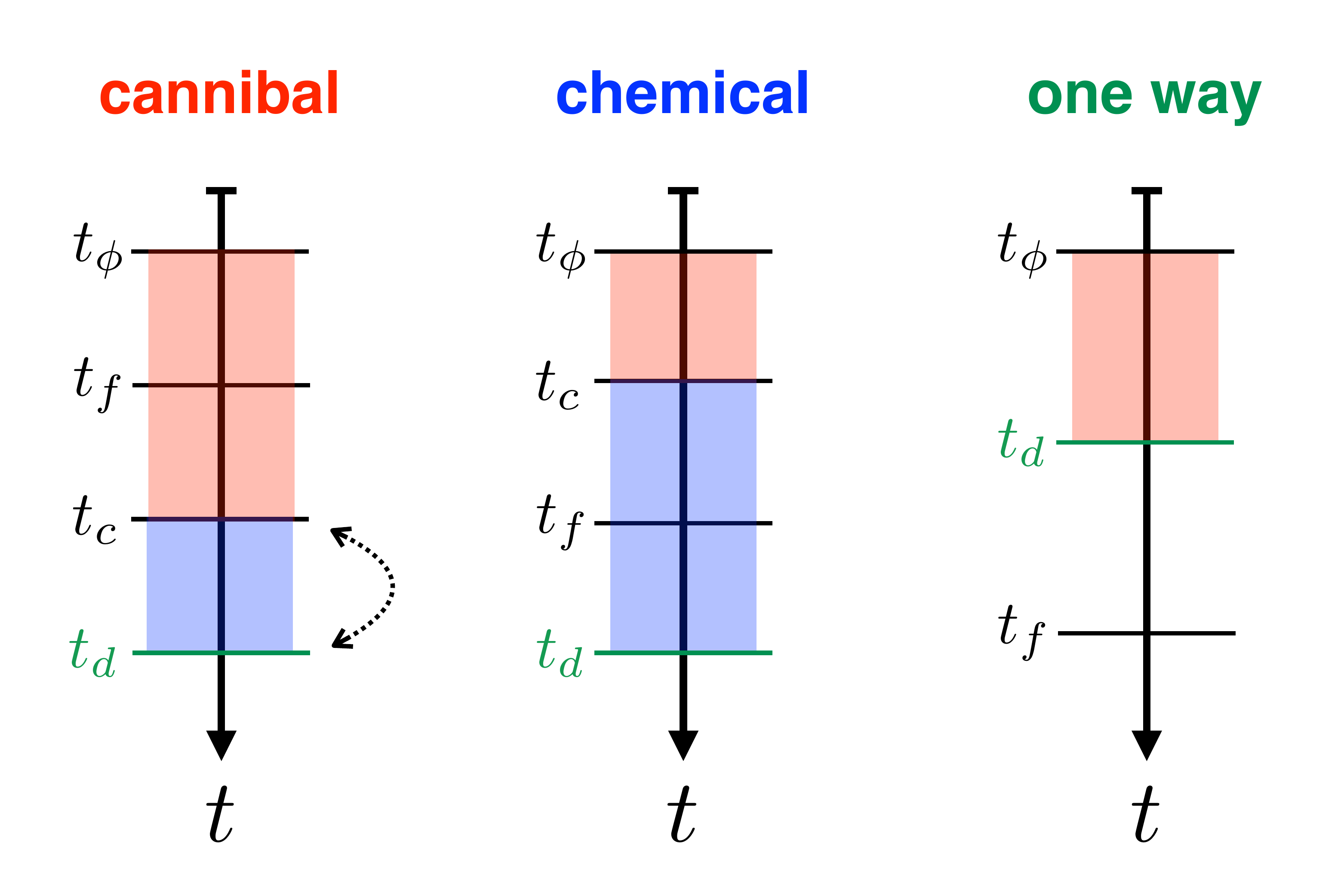}
\end{center}
\vspace{-.3cm}
\caption{\small The three phases of cannibal dark matter: {\it cannibal}, {\it chemical}, and {\it one way}.  The phase depends on the ordering of three timescales: $t_f$, $t_c$, and $t_d$ (which are defined in the text).   The hidden sector undergoes cannibalism in the red shaded region, which begins when $T_d = m_0$ at time $t_\phi$.  Cannibalism ends whichever occurs first: $t_c$ or $t_d$, and either  ordering is possible for the cannibal phase.  The number of hidden particles is conserved in the blue shaded region, because only 2-to-2 interactions are in equilibrium, implying a nonzero chemical potential.
}
\label{fig:schematic}
\end{figure}
%%%%%%%%%%%%%
%%%%%%%%%%%%%
%%%%%%%%%%%%%
%%%%%%%%%%%%%

\begin{figure}[!!!t]
\begin{center}
\includegraphics[width=0.5\textwidth]{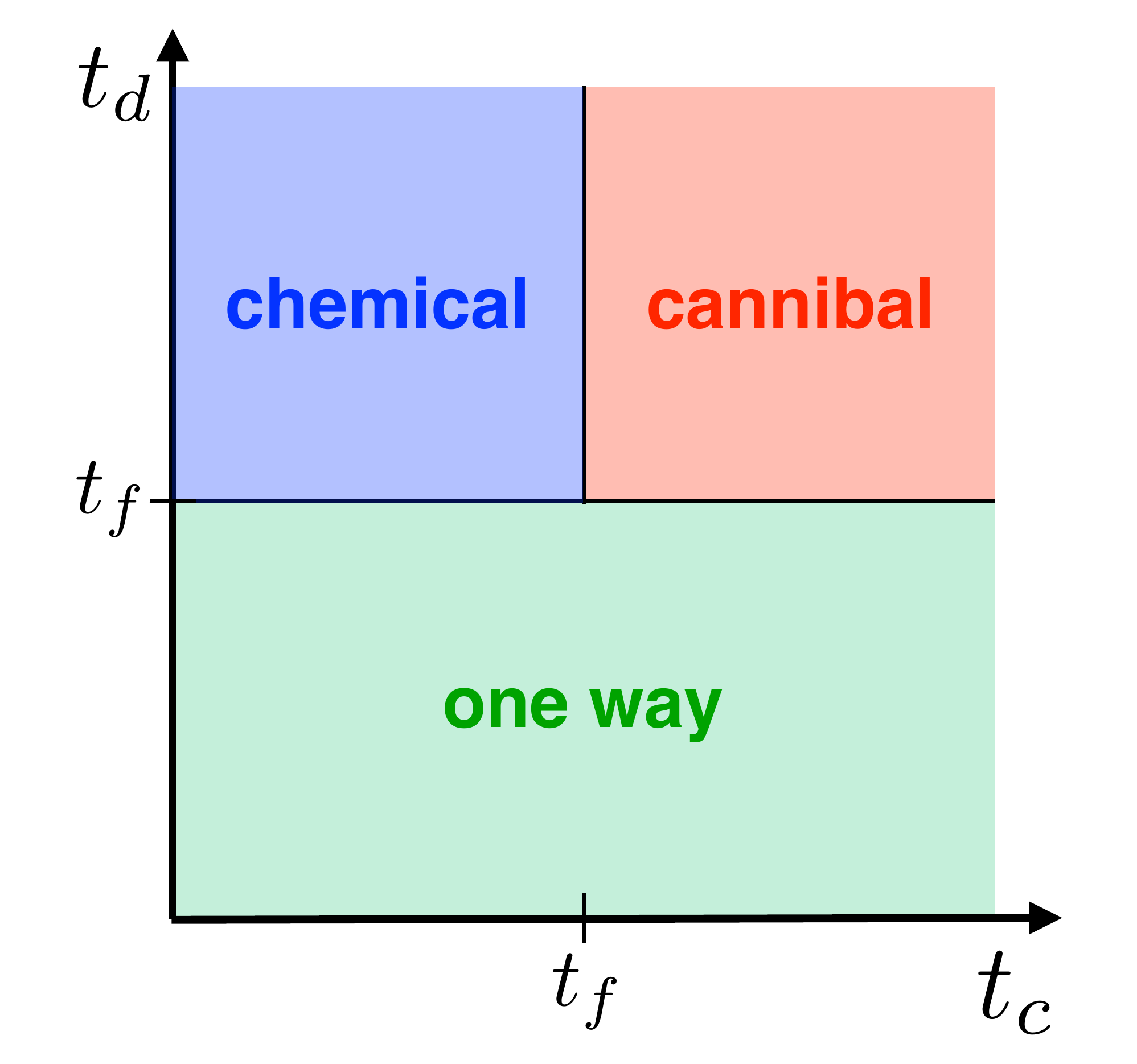}
\end{center}
\vspace{-.3cm}
\caption{\small The phase diagram of cannibal dark matter.  There are three phases that describe the behavior of the hidden sector when DM annihilations freeze-out: {\it cannibal}, {\it chemical}, and {\it one way}.  The phase depends on the ordering of the three timescales: $t_f$, $t_c$, and $t_d$.
The relic density takes a different parametric form within each phase, as shown in Eqs.~\ref{scalingcann}, \ref{scalingchem}, and \ref{scalingdecay}.
}
\label{phases}
\end{figure}

The key point is that different orderings of these timescales lead to different parametric scalings for the DM relic density as a function of the fundamental parameters of the dark sector.  We identify three distinct phases, depending on the order of $t_f$, $t_c$, and $t_d$ (see Fig.~\ref{fig:schematic}).
\begin{itemize}
\item {\bf Cannibal phase} ($t_f\ll t_c,\, t_d$): this is the scenario proposed in Ref.~\cite{cannibal}. The freeze-out of DM number changing interactions takes place while the hidden sector is undergoing cannibalism. The final relic density is exponentially sensitive to the ratio of the LDP and DM masses, $r = m_0 / m_\chi$. The relic density depends on the  $\chi \chi \leftrightarrow \phi \phi$ cross section, $\sigma_2$.
 Assuming the Universe to be radiation dominated at freeze-out,
\be\label{scalingcann}
Y_\chi \propto(m_{\chi}M_P\sigma_2)^{-\frac{1-r}{1-2/3\, r}},
\ee
where $M_P \approx 1.2 \times 10^{19}~\mathrm{GeV}$ is the Planck mass and $Y_{DM} = n_{DM} / s_{SM}$ is the DM yield (see eq.~\ref{Ycann}).   The yield is related to the observed relic density: $\Omega_{DM}/\Omega_{DM}^{\textrm{obs}}=m_{DM}Y_{DM}/(0.4\,{\textrm{eV}})$.  
%%%%
\item {\bf{Chemical phase}} ($t_c\ll t_f\ll t_d$): in this case, at time $t_c$, a chemical potential develops enforcing the conservation of the total number of hidden sector particles.
The DM relic density depends on the size of the number changing cross section, $\sigma_3$, through this chemical potential.  The DM relic density depends inversely on its annihilation cross section, as for a regular WIMP\@. Assuming the Universe to be radiation dominated at freeze-out (see eq.~\ref{Ychem}),
\be\label{scalingchem}
Y_{\chi}\propto \frac{(m_{0}^4 M_P \sigma_3)^{1/4}}{m_{\chi} M_P \sigma_2}.
\ee
Notice the relic abundance is no longer exponentially sensitive to $r$.
%%%%
\item {\bf{One way phase}} ($t_d\ll t_f$ any value of $t_c>t_\phi$):  if the lifetime of the states to which DM is annihilating is shorter than $t_f$, inverse annihilations $\phi \phi \to \chi \chi$ decouple when $\phi$ decays at the time $t_d$.  After $\phi$ decays, forward annihilations $\chi \chi \to \phi \phi$ are still active, but inverse annihilations are suppressed.  The resulting DM yield depends on the width of $\phi$ (see Eq.~\ref{Omegadecay}),
\be\label{scalingdecay}
Y_{\chi}\propto \frac{1}{\Gamma_{\phi}^{1/2}M_P^{3/2}\sigma_2}.
\ee
\end{itemize}
%%%%%%%%%%%%%
%%%%%%%%%%%%%
%%%%%%%%%%%%%
%%%%%%%%%%%%%
\begin{figure}[!!!t]
\begin{center}
~~\includegraphics[width=0.45\textwidth]{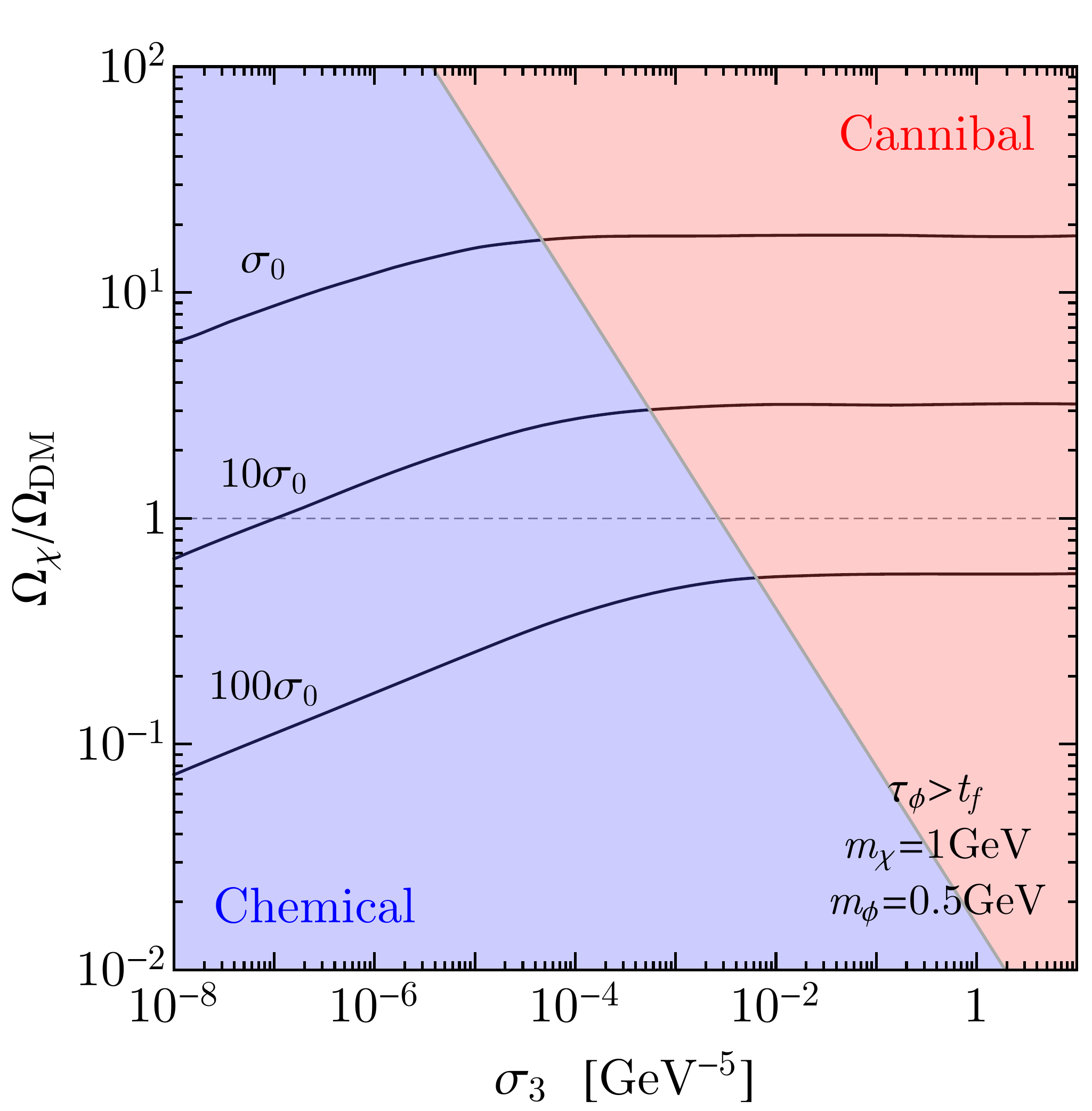}~~~~~\includegraphics[width=0.4639\textwidth]{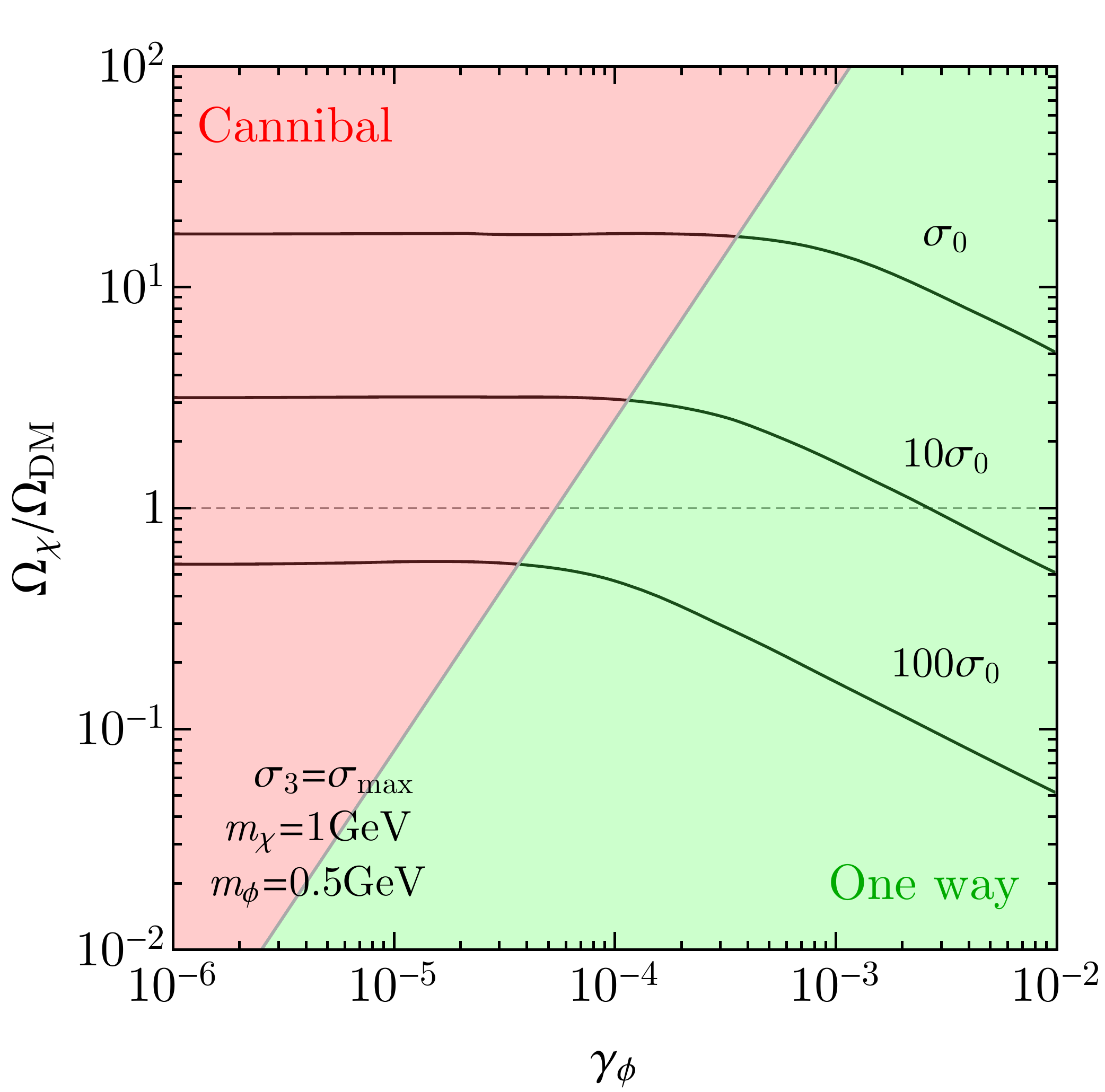}
\end{center}
\vspace{-.3cm}
\caption{ \small
The {\it left} side shows the dark matter relic density as a function of $\sigma_3$ (the cross section of $\phi \phi \phi \rightarrow \phi \phi$), for various values of $\sigma_2$ (the cross section of $\chi \chi \rightarrow \phi \phi$) normalized to the conventional WIMP value $\sigma_0 = 3 \times 10^{-26}~\mathrm{cm}^2/\mathrm{s}$.  Moving from smaller to larger $\sigma_3$, the phase transitions from chemical (blue), where $\Omega_\chi$ increases with $\sigma_3$, to cannibal (red), where $\Omega_\chi$ is independent of $\sigma_3$.  The {\it right} side shows $\Omega_\chi$ as a function of $\gamma_\phi=\Gamma_\phi/H(m_\phi)$.  Moving from smaller to larger $\Gamma_\phi$, the phase transitions from cannibal (red), where $\Omega_\chi$ is independent of $\Gamma_\phi$, to one way (green) where $\Omega_\chi$ decreases with $\Gamma_\chi$. This figure was made using the model of Eqs.~\ref{eq:L} and~\ref{potential}.
}
\label{transition}
\end{figure}
%%%%%%%%%%%%%
%%%%%%%%%%%%%
%%%%%%%%%%%%%
%%%%%%%%%%%%%

These three regimes are displayed in the phase diagram of Fig.~\ref{phases}. Fig.~\ref{transition} shows the behavior of the relic abundance, in terms of the relevant parameters, in the three phases and the transition region between them. In order to draw the curves we use the example model which will be introduced in Section~\ref{sec:GenCo}.

Cannibal DM has distinctive phenomenology, which we explore below.  Most notably, the DM annihilation rate is generically boosted above the standard value of a thermal WIMP, $\sigma_0 \approx 3 \times 10^{26}~\mathrm{cm}^3/\mathrm{s}$.   This is a consequence of the novel parametric form of the DM yield (Eqs.~\ref{scalingcann}, \ref{scalingchem}, and \ref{scalingdecay}) and the requirement that Cannibal DM match the observed DM energy density.  The parametric form of the boost to the cross section is summarized in Tab.~\ref{tab:boost} of Section~\ref{sec:GenCo}.  The boost implies that Cannibal DM is easier to see through indirect detection than standard WIMPs.  Observational constraints on cannibal DM, and future reach, are summarized in Figs.~\ref{figpheno1} and \ref{figpheno2}.  Cannibal DM is constrained by Fermi measurements of $\gamma$-rays, Planck measurements of the Cosmic Microwave Background (CMB), and Lyman-$\alpha$ constraints on warm DM\@.  However, significant parameter space remains allowed and we find that there is promising reach to discover cannibal DM through future CMB measurements.

The rest of this paper is organized as follows. In Section~\ref{sec:Thermo}, we follow the evolution of a decoupled hidden sector as its temperature drops below the mass of the LDP, but when its number changing interactions are still active. In Section~\ref{sec:GenCo}, we introduce the example model that we use to describe the evolution of the DM density during the three phases. In Section~\ref{sec:pheno}, we study possible observational signatures and constraints on the model.  Appendix~\ref{boltzmann} contains a complete description of the system of Boltzmann equations that we use to determine the DM energy density.

%%%%%%%%%%%%%%%%%%%%%%%%%%%%%%%%%%%%%%%%%%%%%%%%%%%%%%%%%%%%%%%%%%%%%%%%%%%%%%%%%%%%%%%%%%%%%%%%%%%%%%%%%%%%%%%%%%%%%%%%%%%%%%%%%%%%%%%%%%%%%%%%%%%%%%%%%%%%%%%%%%%%%%%%%%%%%%%%%%%%%%%%%%%%%%%%%%%%%%%%%%%%%%%%%%%%%%%%%%%%%%%%%%%%%%%%%%%%%%%%%%%%%%%%%%%%%%%%%%%%%%%%%%%%%%%%%%%%%%%%%%%%%%%%%%%%%%%%%%%%%%%%%%%%%%%%%%%%%%%%%%%%%%%%%%%%%%%%%%%%%%%%%%%%%%%%%%%%%%%%%%%%%%%%%%%%%%%%%%%%%%%%%%%%%%%%%%%%%%%%%%%%%%%%%%%%%%%%%%%%%%%%%%%%%%%%%%%%%%%%%%%%%%%%%%%%%%%%%%%%%%%%%%%%%%%%%%%%%%%%%%%%%%%%%%%%%%%%%%%%%%%%%%%%%%%%%%%%%%
\section{Thermodynamics of a Non-Relativistic Hidden Sector}
\label{sec:Thermo}
%%%%%%%%%%%%%%%%%%%%%%%%%%%%%%%%%%%%%%%%%%%%%%%%%%%%%%%%%%%%%%%%%%%%%%%%%%

We now consider the thermodynamic evolution of the hidden sector.  We assume that the hidden sector and the SM are kinetically decoupled from each other, such that they evolve with different temperatures.  In what follows we will refer to SM  and Dark sector quantities with the subscripts $_{\text{SM}}$ and $_{\text{D}}$, respectively.

If both the SM and the hidden sector are separately in thermal equilibrium, the comoving entropies are separately conserved within each sector.  We define the ratio of comoving entropy densities,
\be\label{xidef}
\xi\equiv \frac{s_{SM}}{s_D},
\ee
which is constant throughout the cosmological evolution, while thermal equilibrium is maintained.
For dark sector temperatures $T_D\gg m_0$, the hidden sector  entropy is dominated by relativistic species: $s_d=(2\pi^2/45) g_{*S}^D T_D^3$. This implies that the hidden sector and SM temperatures (apart from a mild dependence due changing $g_{*S}^{1/3}$) are simply proportional to each other~\cite{Feng:2008mu},
\be
T_D=\xi^{-1/3}\left(\frac{g_{*S}^{SM}}{g_{*S}^D}\right)^{1/3}T_{SM},
\ee
where $g_{*S}^{SM}$ and $g_{*S}^D$ are the number of relativistic degrees of freedom in the SM and the hidden sector, respectively.  Note that for the remainder of this paper, we approximate $g_{*S}^{SM,D} \approx g_{*}^{SM,D}$.

Due to the adiabatic expansion of the Universe, the dark temperature $T_D$ will eventually drop  below the mass of the LDP\@. Assuming thermal equilibrium to hold, the phase space distribution of a particle species $X$ will be the non-relativistic Boltzmann distribution $f_X(p)=e^{\mu_X(T)/T}e^{-E(p)/T}$, where we allow the presence of a temperature dependent chemical potential $\mu_X$. It can be shown that the total comoving entropy is still approximately conserved in this regime \cite{Bernstein:1988bw},
\be\label{entropy}
s_D=\sum_X\frac{\rho_X-\mu_X n_X}{T_D}+n_X\approx \frac{m_{0}-\mu_{0}(T_D)}{T_D}n_{0}
\ee
where $X = \chi, \ldots, \phi_0$ sums over dark states and we use the fact that for $T_D\ll m_0$ the entropy is dominated by the lightest species\footnote{\label{foot1}Notice that all particles $\phi_i$ for which $m_{i}-m_{0}\ll T_D$ should be kept in the sum.}. 

If the hidden sector has number changing processes that are still active ({\it i.e.} $T_c\ll m_0$), then the chemical potential vanishes, $\mu_{0}=0$. This condition is easily satisfied.  For example, chemical equilibrium can be maintained by $3\to 2$ reactions involving the LDP,  such as $\phi_0\phi_0\phi_0\to \phi_0\phi_0$ with thermal cross section $\langle\sigma v^2\rangle$.   The equilibrium $\phi_0$ number density is $\bar n_{\phi_0}\equiv n_{\phi_0}^{eq}(T_D,\mu=0)$.   Requiring $\bar n^{2}_{\phi_0}\langle\sigma v^2\rangle>H$ at $T_D=m_{0}$, is equivalent to
\be\label{chemeq32}
\langle\sigma v^2\rangle\gtrsim 10^2\times\frac{\xi^{2/3}g_{*0}^{-4/3}g_{*SM}^{-1/6}}{m_0^4 M_P}\qquad (3\to 2)\,,
\ee
where $g_{*0}$ is the number of degrees of freedom of $\phi_0$ (see footnote~\ref{foot1}). A similar condition holds if chemical equilibrium is maintained by $4\to 2$ reactions $\phi_0\phi_0\phi_0\phi_0\leftrightarrow \phi_0\phi_0$ (or similarly $4 \to 3$ reactions $\phi_0\phi_0\phi_0\phi_0\leftrightarrow \phi_0\phi_0 \phi_0$), in this case
\be\label{chemeq42}
\langle\sigma v^3\rangle\gtrsim 10^3\times\frac{\xi^{2/3}g_{*0}^{-7/3}g_{*SM}^{-1/6}}{m_0^7 M_P} \qquad (4\to 2,3)\,.
\ee

Conservation of comoving entropy implies
\be\label{tempvsa}
s_D \,a^3\propto \left(\frac{T_D}{m_0}\right)^{1/2} e^{-m_0/T_D}\, a^3~~~\Rightarrow~~~ T_D\sim \frac{m_0}{\log a^3}.
\ee
%%%%%%%%%%%%%
%%%%%%%%%%%%%
%%%%%%%%%%%%%
%%%%%%%%%%%%%
\begin{figure}[!!!t]
\begin{center}
\includegraphics[width=1\textwidth]{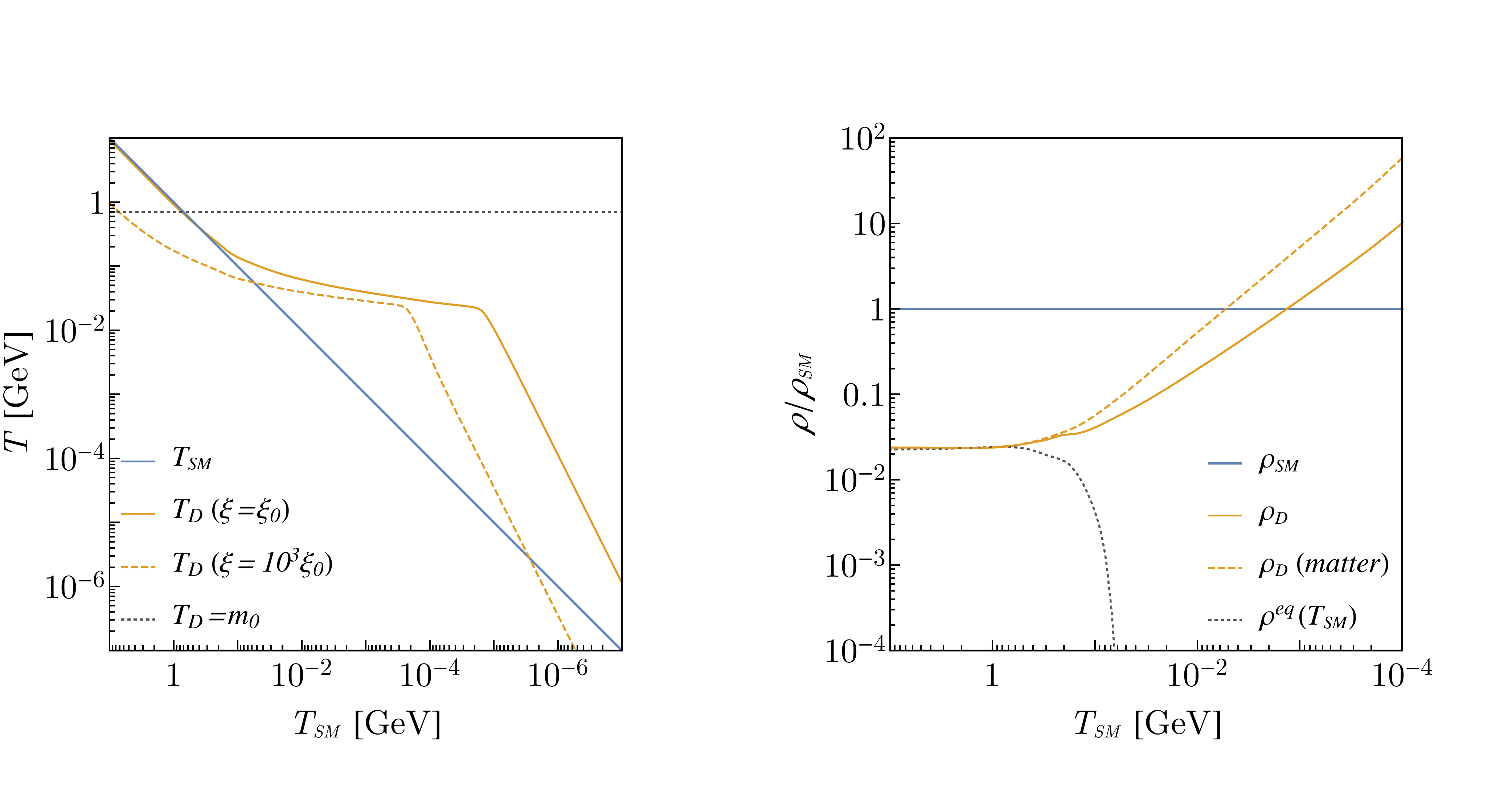}
\end{center}
\vspace{-.3cm}
\caption{\small 
The {\it left} side shows the evolution of the dark temperature, $T_D$, versus the SM temperature, $T_{SM}$, for a cannibalizing sector with LDP mass $m_0 = 0.7~\mathrm{GeV}$.  The solid red (dashed orange) curve corresponds to an initial entropy ratio $\xi = \xi_0~(10^3 \xi_0)$, where $\xi_0 \approx 39$.
On the {\it right} side, the solid red line shows the behavior of the dark energy density over the SM energy density, $\rho_D/\rho_{SM}$, as a function of the SM temperature, for $m_0 = 0.7~\mathrm{GeV}$ and $\xi=\xi_0$.  For comparison, we show  the energy densities of a species in chemical and kinetic equilibrium with the SM, $\rho^{eq}$ (dotted black line, arbitrarily normalized), and of matter, $\rho_D$ (dashed green line).
}
\label{temperature}
\end{figure}
%%%%%%%%%%%%%
%%%%%%%%%%%%%
%%%%%%%%%%%%%
The logarithmic  dependence of the dark sector temperature on the scale factor is the defining feature of \emph{cannibalism}. The hidden sector cannot cool down efficiently as the Universe expands because number changing annihilations are still efficiently converting the rest mass of the light, but non-relativistic, particles into kinetic energy~\cite{Carlson:1992fn}. This behavior is even more striking when one compares the dark sector temperature to the SM one. Using Eqs.~\ref{xidef} and \ref{entropy} we find:
\be
\frac{T_{SM}}{T_D}\simeq 0.52 \,\xi^{1/3}\left(\frac{g_{*0}}{g_{*SM}}\right)^{1/3}\left(\frac{m_0}{T_D}\right)^{5/6}e^{-m_0/3T_D}\,,~~~ T_D<m_0.
\ee
During cannibalism, the SM gets exponentially colder than the dark sector. This behavior is shown in the left panel of Fig.~\ref{temperature}. At the same time, the energy density of the dark sector decreases only logarithmically faster than the energy density of a decoupled pressureless gas ({\it i.e.} matter),
\be
\rho_D\sim \frac{m_0^4}{a^3\log a^3},
\ee
as shown on the right panel of Fig.~\ref{temperature}. Notice that eventually the energy density of the hidden sector will dominate over the SM one. This happens during cannibalism if
\be
T_D\gtrsim \frac{3}{4}\xi\, T_{SM},
\ee
corresponding to $x_0^E\simeq -2.8+\log(\xi^4\, g_{*0}g^{-1}_{*SM})+2.5\log(x_0^E)$, with $x_0^E\equiv m_0/T_D^E$. During cannibalism, as a consequence of Eq.~\ref{tempvsa}, an order one variation in the dark sector temperature corresponds to an exponential increase in the age of the Universe. This holds independently of whether the SM or the dark sector dominates the expansion of the Universe,
\be\label{HubbleTD}
H(T_D)\simeq 
\begin{cases} 
      0.46\, \xi^{2/3}g_{*0}^{2/3}g_{*SM}^{-1/6}\left(T_D/m_0\right)^{1/3}\frac{m_0^2}{M_P}e^{-\frac{2m_0}{3T_D}} & (\rho_{SM} > \rho_D) \\
  0.73\, g_{*0}^{1/2}\left(T_D/m_0\right)^{3/4}\frac{m_0^2}{M_P}e^{-\frac{m_0}{2T_D}}& (\rho_{SM} < \rho_D).  \\
   \end{cases}
\ee

Assuming for the moment that the LDP is stable, as the dark sector cools the rate of number changing interactions will eventually drop below the expansion rate. If number changing processes are dominated by $3\to 2$ annihilations, $\phi_0\phi_0\phi_0\leftrightarrow \phi_0\phi_0$, the condition $\bar n^2_{\phi_0}\langle\sigma v^2\rangle=H$ gives
\be\label{endofcan}
x_{0c}\simeq 
\begin{cases} 
       -14.5+\log (\xi^{-1/2}g_{*SM}^{1/8}g_{*0})+\frac{3}{4}\log (m_0^4 M_P\langle\sigma v^2\rangle) -2\log(x_{0c})& (\rho_{SM} > \rho_D) \\
 -17.4+\log (g_{*0})+\frac{2}{3}\log (m_0^4 M_P\langle\sigma v^2\rangle)-\tfrac{3}{2}\log(x_{0c}) & (\rho_{SM} < \rho_D)  \\
   \end{cases}
\ee
where $x_{0c}=m_0/T_c$. At the time of decoupling the ratio between the SM and the hidden sector temperature is given by
\be
\log \frac{T_{D}}{T_{SM}}\bigg|_{\max}\simeq 
\begin{cases} 
    10.5- \frac{1}{2}\log(\xi g_{*SM}^{-3/4}) +\frac{1}{4}\log (\alpha^3 \textrm{GeV}/m_0)-\frac{3}{2}\log x_{0c}& (\rho_{SM} > \rho_D) \\
9.5- \frac{1}{3}\log(\xi g_{*SM}^{-1}) +\frac{2}{9}\log (\alpha^3 \textrm{GeV}/m_0)-\frac{4}{3}\log x_{0c} & (\rho_{SM} < \rho_D),  \\
   \end{cases}
\ee
where we parametrize $\langle\sigma v^2\rangle\equiv \alpha^3/m_0^5$. After the decoupling of number changing interactions, the total comoving number density is conserved. This, together with entropy conservation, implies that the hidden sector temperature begins to decrease rapidly with the scale factor (see Fig.~\ref{temperature}),
\be\label{tempchem}
\frac{T_D}{T_c}\sim \frac{1}{a^2}.
\ee
Furthermore, a chemical potential for all hidden sector particles is generated,
\be
\mu(T_D)\sim m_0\left(1-\frac{T_D}{T_c}\right),
\ee
enforcing the conservation of the total number of particles in thermal equilibrium. We dub this stage of the thermal evolution of a non-relativistic hidden sector the \emph{chemical era}.   The evolution of the temperature in the chemical era, Eq.~\ref{tempchem},  can be simply understood as coming from the redshift of the velocity of a typical hidden sector particle, $v_\chi\sim 1/a$.  Note that the chemical era is analogous to the evolution of the SM sector after the decoupling of double Compton scattering, after which the number of photons is conserved. After this epoch, the CMB experiences $\mu$-distortions~\cite{Chluba:2011hw}. 

Eventually, as time passes, the various particles in the hidden sector will lose thermal contact with the bath. This happens because the rate of momentum exchange in processes like $\chi \phi \leftrightarrow \chi \phi$ inevitably become slower than the Hubble expansion rate. At some point also $\phi \phi \to \phi \phi$ will decouple and the hidden sector will behave as a gas of non-interacting particles. In general, the thermodynamical treatment given in this section only applies when $\chi \phi \leftrightarrow \chi \phi$ exchanges momentum at a rate larger than Hubble. If this is not the case, the phase space distribution of particles in the thermal bath may deviate substantially from the Boltzmann distribution.

\subsection{LDP Metastability}\label{suddendecay}
Our discussion up to this point has assumed a stable LDP\@. As anticipated in the Introduction, in general this cannot be the case if DM belongs to the hidden sector and $m_{DM}>m_0$. A finite LDP lifetime implies a new timescale $t_d\equiv \Gamma_\phi^{-1}$ to consider. In the following we will assume the LDP (and all other hidden sector particles lighter than the DM) to be unstable and ultimately decay to SM particles.\footnote{Other decay channels can be considered as in Ref.~\cite{cannibal}. For simplicity we will restrict to SM final states.} In order for the decay not to equilibrate the two sectors, we want to ensure the inverse processes $SM\, SM\to \phi$ to be out of equilibrium at $T=T_d$ when $\phi$ decays.  Imposing $\bar n_\gamma(T_{d\,SM})^2 \Gamma_{SM\,SM\to \phi} \ll  n_{\phi_0}(T_d) \Gamma_\phi$ and using detailed balance to write $\bar n_\gamma(T_{d\,SM})^2 \Gamma_{SM\,SM\to \phi}=  \bar n_{\phi_0}(T_{d\,SM}) \Gamma_\phi$ we obtain
\be\label{outofinverseeq}
\frac{\bar n_{\phi_0}(T_{d\,SM})  }{n_{\phi_0}(T_d)} \ll 1,
\ee
where $T_{d\,SM}$ is the SM temperature at time $t_d$.
If the two sectors start with similar temperature above every mass thresholds Eq.~\ref{outofinverseeq} implies the familiar condition
\be\label{outofeq}
\Gamma_\phi \ll H(T_{SM}=m_\phi).
\ee
In general however, if the two sectors have the same temperature at early times, Eq.~\ref{outofeq} can be realized in two ways: $n_\phi$ can be the equilibrium distribution but $T_d\gg T_{d\,SM}$ as it is the case after a epoch of cannibalism; $T_d\approx T_{d\,SM}$ but $n_\phi\gg \bar n_\phi$ because of a chemical potential. Either one of this possibilities will be realized in the following.

As soon as the LDP starts to decay, all reactions involving it begin to decouple, simply because its density starts to decrease exponentially. In particular all processes like $\phi_0\phi_0\leftrightarrow \phi_0\phi_0$ and $\chi \phi_0\leftrightarrow \chi_i \phi_0$ will no longer be able to keep the system in thermal equilibrium: the various assumptions made in the previous section about the phase space distribution of particles and entropy conservation in the hidden sector begin to fail. A correct treatment of the thermal evolution from this point on would require us to solve a full unintegrated Boltzmann equation.  This goes beyond the scope of this work. Sticking to the more modest goal of approximately calculating the DM relic abundance, we will describe the evolution of the system after $t_d$ using a \emph{sudden decay approximation} in which we set $n_{LDP}=0$ at $t=t_d$. We will furthermore assume that all particles, such as $\chi$, which were kept in kinetic equilibrium with the thermal bath by elastic scattering $\chi \phi_0\leftrightarrow \chi \phi_0$ will keep a Boltzmann phase distribution with a temperature redshifting as $T\propto 1/a^2$.

In this sudden decay approximation, the SM plasma is instantaneously reheated by the LDP decay products. If the energy density of the LDP dominates over the energy density of the SM plasma, by energy conservation, the decay of the LDP will reheat the SM to a larger temperature $T_{RH}$ \cite{Kolb:1990vq},
\be\label{TRH}
T_{RH}\equiv\left(\frac{45}{16\pi^3 g_{*SM}}\right)^{1/4}\sqrt{\Gamma_\phi M_P}\simeq 0.55\,  g_{*SM}^{-1/4}\sqrt{\Gamma_\phi M_P},
\ee
with a corresponding increase of the SM entropy.

Notice that if $t_d<t_c$, cannibalism ends because of the LDP decay. We will discuss the additional implications of $t_d<t_f$ in the next section.
%%%%%%%%%%%%%%%%%%%%%%%%%%%%%%%%%%%%%%%%%%%%%%%%%%%%%%%%%%%%%%%%%%%%%%%%%%%%%%%%%%%%%%%%%%%%%%%%%%%%%%%%%%%%%%%%%%%%%%%%%%%%%%%%%%%%%%%%%%%%%%%%%%%%%%%%%%%%%%%%%%%%%%%%%%%%%%%%%%%%%%%%%%%%%%%%%%%%%%%%%%%%%%%%%%%%%%%%%%%%%%%%%%%%%%%%%%%%%%%%%%%%%%%%%%%%%%%%%%%%%%%%%%%%%%%%%%%%%%%%%%%%%%%%%%%%%%%%%%%%%%%%%%%%%%%%%%%%%%%%%%%%%%%%%%%%%%%%%%%%%%%%%%%%%%%%%%%%%%%%%%%%%%%%%%%%%%%%%%%%%%%%%%%%%%%%%%%%%%%%%%%%%%%%%%%%%%%%%%%%%%%%%%%%%%%%%%%%%%%%%%%%%%%%%%%%%%%%%%%%%%%%%%%%%%%%%%%%%%%%%%%%%%%%%%%%%%%%%%%%%%%%%%%%%%%%%%%%%%
\section{Phases of Hidden Freeze-out}
\label{sec:GenCo}
\begin{figure}[!!!t]
\begin{center}
\includegraphics[width=0.55\textwidth]{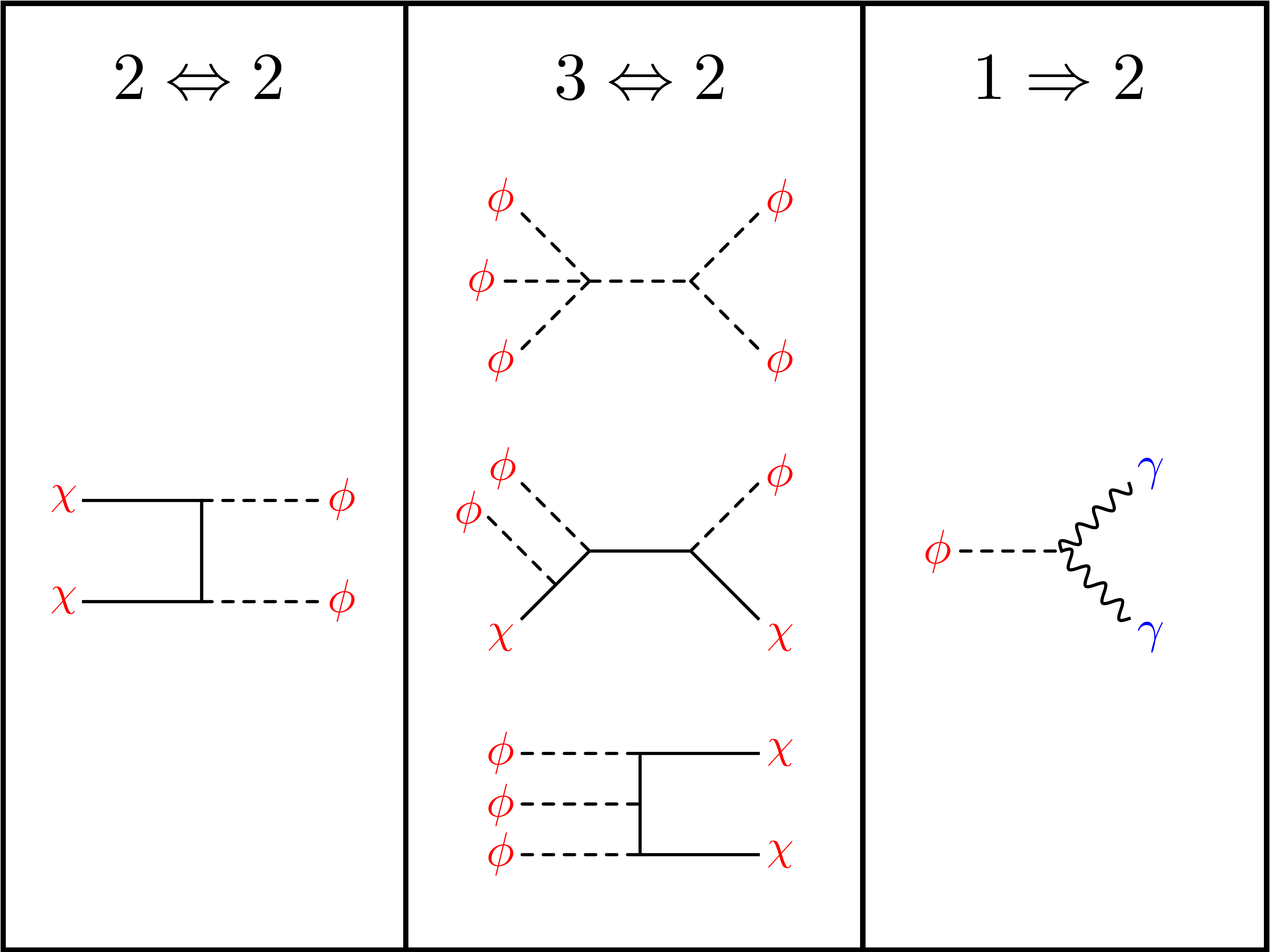}
\end{center}
\vspace{-.3cm}
\caption{\small 
Example Feynman diagrams corresponding to the key processes for determining the $\chi$ relic density from the model of Eqs.~\ref{eq:L} and \ref{potential}.  DM freezes out when $\chi \chi \rightarrow \phi \phi$, shown to the {\it left}, decouples.  The number of hidden particles is conserved after all of the $3 \leftrightarrow 2$ processes shown in the {\it middle} decouple after time $t_c$.  The LDP can decay to the SM through the process $\phi \rightarrow \gamma \gamma$, shown to the {\it right}.
}
\label{fig:feynman}
\end{figure}

%%%%%%%%%%%%%%%%%%%%%%%%%%%%%%%%%%%%%%%%%%%%%%%%%%%%%%%%%%%%%%%%%%%%%%%%%%
In this section we study DM freeze-out in the thermal background of a non-relativistic hidden sector, as discussed in the previous section. In order to simplify the discussion, we introduce a specific model in which all the features discussed above are realized. The hidden sector we consider is composed of a single Majorana fermion, $\chi$, which plays the role of the DM, and a real scalar singlet, $\phi$, playing the role of the LDP\@. The interactions between these degrees of freedom are described by the most general renormalizable Lagrangian\cite{cannibal},
\be \label{eq:L}
\mathcal L=\frac{1}{2}\partial^\mu\phi\partial_\mu\phi-V(\phi)+i\chi^\dagger\bar\sigma^\mu\partial_\mu \chi-\frac{m_\chi}{2}\chi\chi-\frac{y+i y_5}{2}\phi\chi\chi+{\textrm{h.c.}},
\ee
with
\be\label{potential}
V(\phi)=\frac{m_\phi^2}{2}\phi^2+\frac{A}{3!}\phi^3+\frac{\lambda}{4!}\phi^4.
\ee
In order to realize the kinematics of the previous section we require $m_\chi> m_\phi$ and define $r\equiv m_\phi/m_\chi$. We also assume that the parameters in the $\phi$ potential are such that $\phi$ gets no vacuum expectation value. It is possible to reduce the number of free parameters in Eq.~\ref{potential} by requiring $A$ to arise from the spontaneous breaking of a $Z_2$ symmetry, $\varphi \leftrightarrow - \varphi$,
\be\label{potential1}
V=-\frac{m_\varphi^2}{2}\varphi^2+\frac{\lambda}{4!}\varphi^4~~~\Rightarrow ~~~\varphi = \frac{m_\varphi}{\sqrt \lambda} + \phi,~~~m_\phi=\sqrt 2 m_\varphi,~~~{\textrm{and}}~~~ A=\sqrt{3\lambda}m_\phi.
\ee
All parameters in the Lagrangian are real. We will furthermore assume the existence of a coupling 
\be
\mathcal L\supset \frac{\phi}{\Lambda}\mathcal O_{SM}, 
\ee
where $\mathcal O_{SM}$ is some dimension 4 operator made of $SM$ fields, mediating $\phi$ decay with a typical width $\Gamma_\phi\sim m_\phi^3/\Lambda^2$. In the following we take $\mathcal O_{SM}=F_{\mu\nu}^2$ or $F_{\mu\nu}\tilde F_{\mu\nu}$, mediating $\phi$ decay to photons.  Example Feynman diagrams that generate the cosmologically relevant processes are shown in Fig.~\ref{fig:feynman}.

The precise evolution of the DM number density is obtained by solving a set of coupled Boltzmann equations. We are interested in  the evolution of 4 variables: $T_{SM}$, $T_D$, $n_\phi$ and $n_\chi$, as functions of the scale factor $a$. As explained in the previous section, $\Gamma_\phi\neq 0$ cannot be treated exactly. For this reason, in our numerical analysis, we consider two different Boltzmann systems depending on whether $a<a_d$ or $a>a_d$, where $a_d$ is the scale factor at which $H(a_d)=\Gamma_\phi$ ({\it i.e.} at time $t_d$). For $a<a_d$, the dark sector is in thermal equilibrium and the $\chi$ and $\phi$ number densities evolve according to two Boltzmann equations,
\begin{align}\label{boltznumb}
a^{-3}\frac{d (n_\phi a^3)}{dt}&=\mathcal K_{\phi}(\{\langle \sigma_i\rangle\}, T_D,n_\phi, n_\chi),\\
a^{-3}\frac{d (n_\chi a^3)}{dt}&=\mathcal K_{\chi}(\{\langle \sigma_i\rangle\}, T_D,n_\phi, n_\chi).
\end{align}
The set $\{\langle \sigma_i\rangle\}$ includes all relevant cross sections, where the specific form of the kernels $\mathcal K_{\phi,\chi}$ are described in Appendix~\ref{boltzmann}\@. Solving these two Boltzmann equations requires knowledge of the scale factor dependence of $T_D$ and $T_{SM}$. Two additional equations are obtained imposing the separate conservation of the comoving entropy densities,
\be\label{boltztemp}
\frac{d (s_\phi a^3+s_\chi a^3) }{dt}=0~~~{\textrm{and}}~~~ \frac{d (s_{SM} a^3)}{dt}=0,
\ee
where $s_\phi$ and $s_\chi$ are given by the entropy formula in Eq.~\ref{entropy}. These 4 equations constitute a closed system that can be solved numerically for $a<a_d$. At $a=a_d$, the thermal equilibrium assumption that goes into Eqs.~\ref{boltznumb} and \ref{boltztemp} is no longer valid. In order to estimate the final DM density, we modify the Boltzmann system by using the following sudden decay approximation:
\begin{itemize}
\item We set $n_\phi(a)=0$ for $a>a_d$, dropping the $n_\phi$ Boltzmann equation and using $n_\chi(a_d^+)=n_\chi(a_d^-)$ as a boundary condition for $n_\chi$.\footnote{Notice that this assumption can be justified as long as $\chi\chi\to\chi\chi$ exchanges momentum efficiently.}
\item For $a>a_d$, we set $T_D(a)/T_D(a^-_d)=a_d^2/a^2$ and we drop the equation enforcing the conservation of the comoving hidden sector entropy.
\item As a boundary condition for the SM entropy equation, we use $s_{SM}(a_d^+)=2\pi^2/45 g_{*SM}T_{RH}^3$, where $T_{RH}$ is defined in Eq.~\ref{TRH}.
\end{itemize}

The exact form of the system of differential equations that we solve to obtain our numerical results is shown in Appendix~\ref{boltzmann}\@. In the following three sections, we provide an analytic understanding of the dependence of the DM relic abundance on the various parameters of the model. The analytic formulae we are going to discuss work well when the three timescales $t_d$, $t_f$ and $t_c$ are well separated and $g_{*SM}$ can be approximated as constant during DM freeze-out (requiring that DM not freeze-out during the QCD phase transition, for example).

Starting at high temperatures, $T_D\gg m_\chi$, the number of relativistic hidden sector degrees of freedom is $g_{*D}=1+7/8\times 2=2.75$. If the SM and the hidden sector were in thermal contact at very high temperatures, for instance above all the SM thresholds, this fixes the initial condition for $\xi$,
\be
\xi=\frac{g_{*SM}}{g_{*D}}\equiv \xi_0\approx 39.
\ee
Different values for $\xi$ are of course possible if thermal equilibrium between the two sectors was never attained.\footnote{There is an upper bound on the value of $\xi$ in the presence of a non-vanishing decay rate. $SM SM\to \phi$ processes dump entropy in the hidden sector until $T_{SM}\sim m_\phi$. This implies a minimal $\xi$ of order $m_\phi^2/(M_P\Gamma_\phi)$ \cite{lawrence}.}
We allow generic $\xi$ to discuss various semi-analytical estimates in the following, but fix $\xi=\xi_0$ in all plots (unless otherwise noted). 

As time evolves, the hidden sector temperature eventually drops below $m_\phi$. In the following, we assume that this transition happens while the hidden sector is still in chemical equilibrium.  A condition analogous to Eq.~\ref{chemeq32} therefore applies, such that the total number of hidden sector particles is not conserved.  If this is the case, as we explained in the previous section, the hidden sector goes through an epoch of cannibalism.  The number changing processes we consider in the following are $3\to 2$ reactions involving $\chi$ and $\phi$, for instance $\phi\phi\phi\leftrightarrow\phi\phi$, $\chi\phi\phi\leftrightarrow\chi\phi$, $\phi\phi\phi  \leftrightarrow \chi\chi,\,\ldots$.  Typically, $\phi\phi\phi\leftrightarrow\phi\phi$ is the most efficient of these processes, and its decoupling sets the departure from chemical equilibrium.
Compared to the rate of $\phi\phi\phi\leftrightarrow\phi\phi$, the rates of $3\to 2$ reactions involving $\chi$ are proportional to $|y|^2$ and Boltzmann suppressed by a factor $e^{-(m_\chi-m_\phi)/T}$.\footnote{This is true for all processes except $\phi\phi\phi\to\chi\chi$ when $3m_\phi>2m_\chi$. In this case, while it is still true that the rate is proportional to $|y|^2$, there is no additional Boltzmann suppression.} For these reasons, their effects are always subleading or negligible. While we include all the processes in our numerical calculations (through the functions $\mathcal K_\phi$ and $\mathcal K_\chi$ in Eq.~\ref{boltznumb}), our analytical discussion focuses on $\phi\phi\phi\leftrightarrow\phi\phi$. The thermally averaged cross section for this process, using the potential in Eq.~\ref{potential1}, is
\be
\langle\sigma_{\phi\phi\phi\to\phi\phi} v^2\rangle\equiv\sigma_3=\frac{25\sqrt{5}\,\lambda^3}{3072\,\pi m_\phi^5} + \mathcal{O}(T_D / m_\phi)
\ee
and its perturbative upper limit $\sigma_3^{max}$ is obtained setting $\lambda=16\pi^2$. This cross section can be used to calculate the temperature $T_c$ (see Eq.~\ref{endofcan}) at which cannibalism ends and the chemical era begins.

The final dark matter abundance depends on the rate of $\chi\chi\to\phi\phi$ annihilations. Depending on the complex phase of the coupling $y$, this $2\to2$ annihilation cross section is either $s$-wave or $p$-wave. The first case is realized if either $y_5A\neq0$ or $y y_5\neq 0$. For $s$-wave annihilations, taking $m_\phi=0$ for simplicity,
\be
\langle\sigma^{(s)}_{\chi\chi\to\phi\phi} v\rangle\equiv \sigma_{2}^{(s)}=\frac{y_5^2\left(y+\tfrac{A}{8 m_\chi}\right)^2}{16\pi m_\chi^2} + \mathcal{O}(T_D / m_\chi).
\ee
In all other cases the cross section is $p$-wave and, again in the $m_\phi=0$ limit and assuming $y_5=0$ for simplicity,
\be
\langle\sigma^{(p)}_{\chi\chi\to\phi\phi} v\rangle\equiv \sigma_{2}^{(p)}=\frac{y^2\left(y^2+\tfrac{10}{9}\tfrac{y A}{m_\chi}+\tfrac{1}{48}\tfrac{A^2}{m^2_\chi}\right)}{128\pi m_\chi^2}\times \frac{T_D}{m_\chi} + \mathcal{O}(T_D^2 / m_\chi^2)
\ee
The full expressions, including nonzero $m_\phi$ and $y_5$, are reported in Appendix \ref{boltzmann}\@. 

Neglecting the finite width of $\phi$, freeze-out of the DM number density roughly occurs at a dark temperature $T_f$ defined by
\be\label{freezeout}
\bar n_\chi\sigma_2= H.
\ee
Depending on whether $T_c<T_f$ or $T_f<T_c$, freeze-out  occurs in the cannibal or chemical phase, respectively, changing the expectation for the relic abundance. The dependence of the final abundance on the model parameters will be discussed in the next two sections. Notice that in both cases, when Eq.~\ref{freezeout} is satisfied, both reactions $\chi\chi\to \phi\phi$ and $\phi\phi\to\chi\chi$ have become inefficient: it is rare for two $\chi$s to find each other to annihilate because of the Boltzmann suppression of their distribution and at the same time even though $\phi$s are exponentially more numerous, the temperature is too low to allow them to annihilate into the heavier $\chi$s. The final comoving number density is then roughly fixed by the equilibrium value at $T_f$.

A completely different evolution takes place if $\phi$ decays before the $\chi$ abundance freezes out ({\it i.e.} the one way phase).  At the dark sector decay temperature $T_d$, the rate of $\chi\chi\to\phi\phi$ annihilations is still larger than Hubble, while the inverse process, $\phi\phi\to\chi\chi$, decouples. $\chi$ continues to annihilate and, as we show in Section \ref{subsec:decay}, the final DM abundance is  smaller than the $t_f<t_d$ case. 

\begin{figure}[!!!t]
\begin{center}
\includegraphics[width=0.55\textwidth]{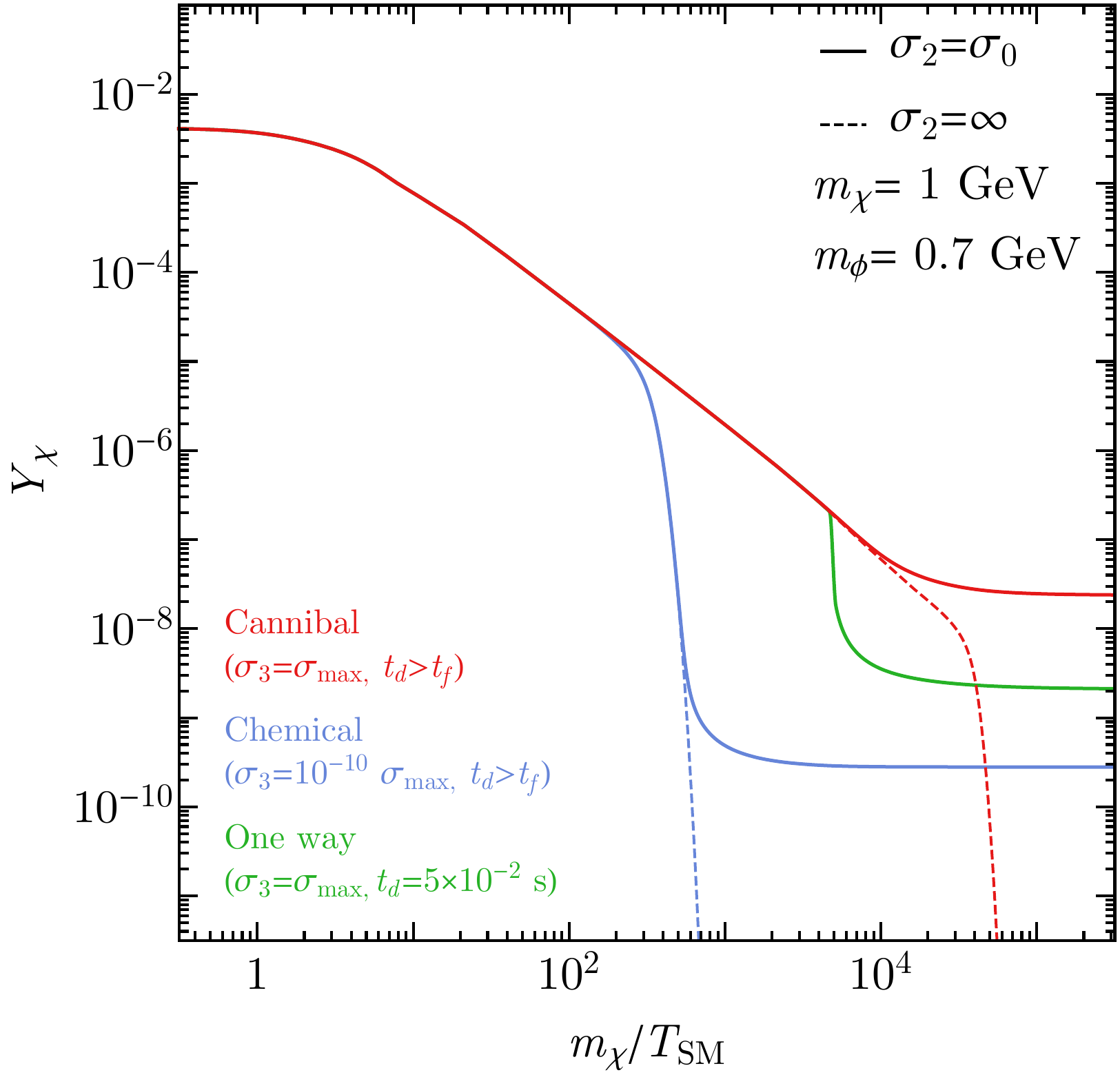}
\end{center}
\vspace{-.3cm}
\caption{\small Examples of the DM yield, $Y_\chi = n_\chi / s_{SM}$, verses inverse temperature, $m_\chi / T_{SM}$, for each of the phases of DM freeze-out: cannibal, chemical, and one way.  The solid curves show the behavior of the yield with the $\chi \chi \rightarrow \phi \phi$ cross section, $\sigma_2$, fixed to the standard WIMP value $\sigma_0 = 3 \times 10^{-26}~\mathrm{cm}^3/\mathrm{s}$.  For comparison, the dashed curves show the yield if $\sigma_2$ is taken large enough that $\chi$ annihilations never decouple.
During cannibalism, the equilibrium DM abundance scales as $Y_\chi^{eq} \propto T_{SM}^{3/r-3}$ with $r = m_\phi / m_\chi$.  For the cannibal phase, $Y_\chi$ freezes out at $t_f$.  For the chemical phase, cannibalism ends at $t_c$ and then $Y_\chi$ drops exponentially with $T_{SM}$ until it freezes out at $t_f$.  For the one way phase, $Y_\chi$ departs from equilibrium when $\phi$ decays and then $Y_\chi$ immediately freezes out.
}
\label{fig:yield}
\end{figure}

The yield in the cannibal, chemical, and one-way phase is shown in Fig.~\ref{fig:yield} for a specific choice of parameters.

%%%%%%%%%%%%%%%%%%%%%%%%%%%%%%%%%%%%%%%%%%%%%%%%%%%%%%%%%%%%%%%%%%%%%%%%%%%%%%%%%%%%%%%%%%%%%%%%%%%%%%%%%%%%%%%%%%%%%%%%%%%%%%%%%%%%%%%%%%%%%%%%%%%%%%%%%%%%%%%%%%%%%%%%%%%%%%%%%%%%%%%%%%%%%%%%%%%%%%%%%%%%%%%%%%%%%%%%%%%%%%%%%%%%%%%%%%%%%%%%%%%%%%%%%%%%%%%%%%%%%%%%%%%%%%%%%%%%%%%%%%%%%%%%%%%%%%%%%%%%%%%%%%%%%%%%%%%%%%%%%%%%%%%%%%%%%%%%%%%%%%%%%%%%%%%%%%%%%%%%%%%%%%%%%%%%%%%%%%%%%%%%%%%%%%%%%%%%%%%%%%%%%%%%%%%%%%%%%%%%%%%%%%%%%%%%%%%%%%%%%%%%%%%%%%%%%%%%%%%%%%%%%%%%%%%%%%%%%%%%%%%%%%%%%%%%%%%%%%%%%%%%%%%%%%%%%%%%%%
\subsection{Cannibal Dark Matter Review}
\label{subsec:cannibal}
%%%%%%%%%%%%%%%%%%%%%%%%%%%%%%%%%%%%%%%%%%%%%%%%%%%%%%%%%%%%%%%%%%%%%%%%%%
%%%%%%%%%%%%%
%%%%%%%%%%%%%
%%%%%%%%%%%%%
%%%%%%%%%%%%%
\begin{figure}[!!!t]
\begin{center}
\includegraphics[width=0.9\textwidth]{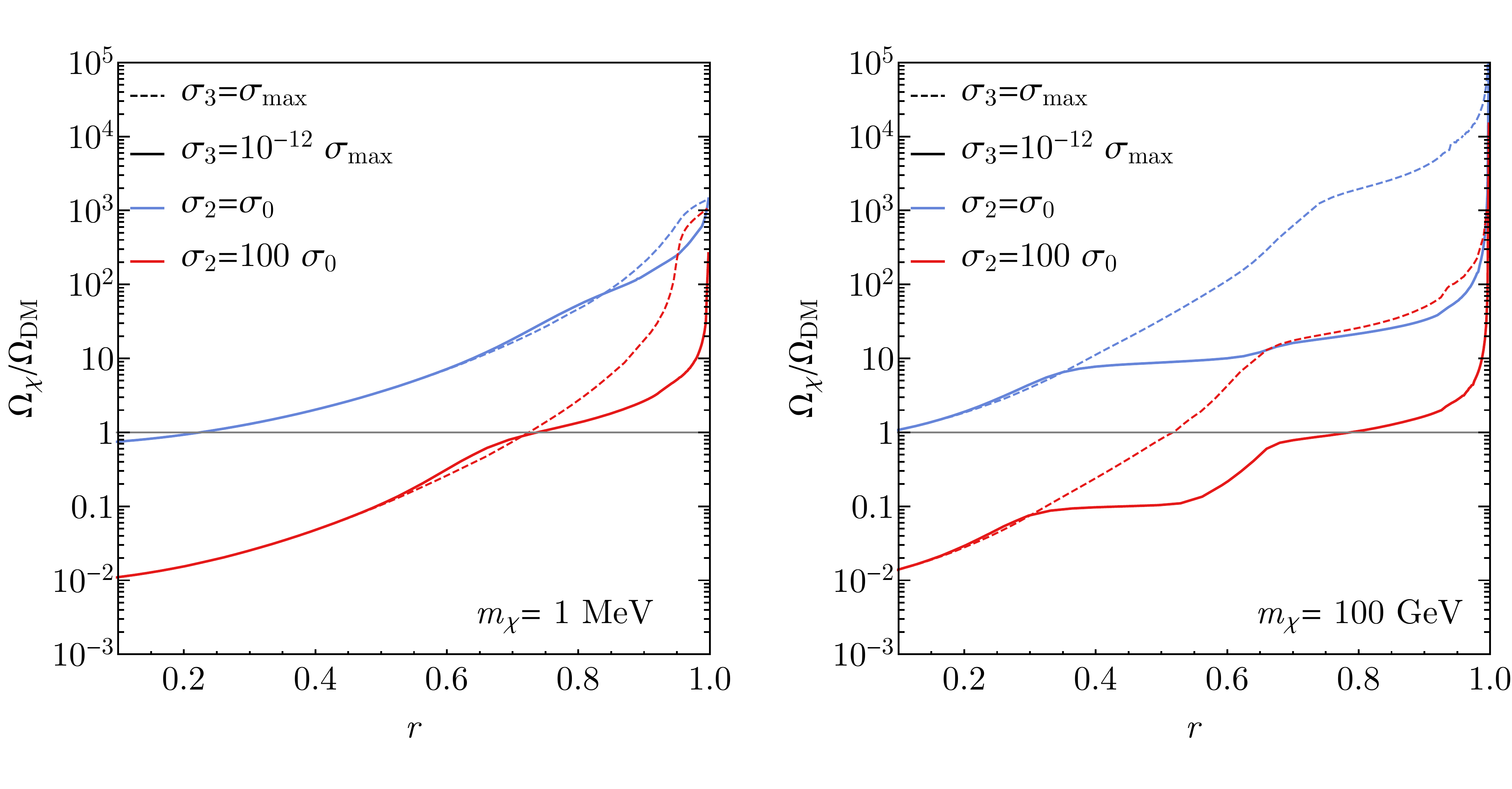}
\end{center}
\vspace{-.3cm}
\caption{\small  Behavior of $\Omega_\chi/\Omega_{DM}$ as a function of $r$, for various choices of $m_\chi$, $\sigma_2$ and $\sigma_3$. The $\phi$ decay width is such that $t_d\gg t_f, t_c$, but such that $D\sim O(1)$.  The three regimes discussed in the text are visible in the plot and are characterized by different slopes of the various curves. On the leftmost portion of each curve, freeze-out happens in the cannibal phase.  Depending on the various parameters, a certain $r_*$ is reached such that freeze-out begins to happen in the chemical phase. For $r$ very close to 1, $2\to 2$ freeze-out happens when $1-r<x^{-1}_f$ and the comoving number density of $\chi$ is equal to that of $\phi$ at the freeze-out of $3\to 2$ annihilations.  In this limit, one recovers the behavior of Ref.~\cite{Carlson:1992fn}, as described in the text.
}
\label{Omegar}
\end{figure}
%%%%%%%%%%%%%
%%%%%%%%%%%%%
%%%%%%%%%%%%%
%%%%%%%%%%%%%
If $T_c\ll T_f$ (and $T_d$ is smaller than both), then the final DM relic abundance is well described by a single Boltzmann equation \cite{cannibal}:
\be\label{boltzmanncannibal}
a^{-3}\frac{d (n_\chi a^3)}{d\log a}=-\frac{\sigma_2}{H}(n_\chi^2-\bar n_\chi^2),
\ee
where $\sigma_2$ is the thermal average of the $\chi\chi\to\phi\phi$ cross section, which we assume to be $s-$wave in the following, for simplicity. To this equation, we add equations describing conservation of  entropy (Eq.~\ref{boltztemp}) giving the scale factor evolution of $T_D$ and $T_{SM}$. 
Note that Eq.~\ref{boltzmanncannibal} is the usual freeze-out equation for a WIMP\@.  The non-standard behavior is the dependence of $T_D$, and hence $\bar n_{\chi}$and $H$, on the scale factor.
The comoving relic abundance of $\chi$ is approximately equal to the equilibrium number density at freeze-out,
\be
\bar n_\chi\sigma_2= H.
\ee
Using the equilibrium distribution for $\chi$,
\be\label{nchieq}
\bar n_\chi(T_D)=2\times \left(\frac{m_\chi T_D}{2\pi}\right)^{3/2} e^{-m_\chi/T_D},
\ee
and the expression for Hubble as a function of $T_D$ in Eq.~\ref{HubbleTD}, we obtain the expression for the temperature at freeze-out ($x_f=m_\chi/T_f$):
\be\label{cannFO}
(1-\alpha r) x_f\simeq \log (h(r) m_\chi M_P\sigma_2)-\beta \log (x_f),
\ee
where
\be\label{cannh}
h(r)\simeq r^{-1/2-\beta}\times\begin{cases} 
    0.3\, \xi^{-2/3}g_{*SM}^{1/6}& (\rho_{SM} > \rho_D)\\
   0.2& (\rho_{SM} < \rho_D),
   \end{cases}
\ee
and $\alpha=2/3$, $1/2$ and $\beta=7/6$, $3/4$ depending on whether the SM or the hidden sector dominate the Hubble expansion at freeze-out. In order to calculate the DM yield at freeze-out, $Y^f_\chi=m_\chi n_\chi/s_{SM}$, we plug Eq.~\ref{cannFO} into \ref{nchieq}, obtaining
\be\label{Ycann}
Y^f_\chi\simeq\frac{1}{\xi}\frac{2}{r^{5/2}x_f}\times\left[\frac{ x_f^\beta}{h(r)}\right]^{\frac{1-r}{1-\alpha r}}(m_\chi M_P\sigma_2)^{-\frac{1-r}{1-\alpha r}}.
\ee
These expressions hold only in the limit in which $T_f\gg T_c$, with $T_c$ as in Eq.~\ref{endofcan}. In particular, there is a maximal $r$ beyond which the condition of Eq.~\ref{endofcan} is no longer satisfied.  The ratio between the rate of $3\to2$ reactions and that of $2\to2$ is exponentially sensitive to $r$,
\be
\frac{\bar n_\phi^2 \sigma_3}{\bar n_\chi\sigma_2}\propto e^{(m_\chi-2m_\phi)/T_D},
\ee 
and for $r>1/2$ a large hierarchy between $\sigma_3$\ and $\sigma_2$ is needed for $T_f\gg T_c$ to be satisfied.

The final dark matter abundance is given by
\be\label{Omegacann}
\frac{\Omega_{\chi}}{\Omega^{\textrm{obs}}_{\rm DM}}\approx 
\begin{cases} 
      0.3\, \frac{x_f}{g_{*SM}^{1/2}} \frac{\sigma_0}{\sigma_2} \frac{T_f}{T_{fSM}}\times \frac{1}{D} & (\rho_{\rm SM} > \rho_D) \\
   0.3\,  \frac{x_f}{g_{*SM}^{1/2}}  \frac{\sigma_0}{\sigma_2} \frac{T_f^{3/2}}{\xi^{1/2} T_{fSM}^{3/2}}\times\frac{1}{D}& (\rho_{\rm SM} < \rho_D),  \\
   \end{cases}
\ee
where $\sigma_0=3\times 10^{-26}\,{\textrm{cm}}^3{\textrm{s}}^{-1}$ is the typical WIMP freeze-out cross section, and $D$ is a dilution factor accounting for the entropy injection due to $\phi$ decay. All temperatures are evaluated at freeze-out of $2\to 2$. $D$ is different from 1 only if $\phi$ is dominating the energy density of the Universe at the time it decays. If this is the case $D$ is calculated using the sudden decay approximation described in section \ref{suddendecay} and approximately $D\approx T^{SM}_E/T_{RH}$, where $T_E$ is the SM temperature when $\phi$ starts to dominate the energy density of the universe. Eq.~\ref{Omegacann} shows that (neglecting $D$ for simplicity) the thermal cross section needed to obtain the right relic abundance is boosted by a factor corresponding to the ratio of temperature between the dark sector and the SM at freeze-out. The behavior of $\Omega_{\chi}/\Omega_{\rm DM}$, as a function of $r$, is shown in Fig.~\ref{Omegar}.

%%%%%%%%%%%%%%%%%%%%%%%%%%%%%%%%%%%%%%%%%%%%%%%%%%%%%%%%%%%%%%%%%%%%%%%%%%%%%%%%%%%%%%%%%%%%%%%%%%%%%%%%%%%%%%%%%%%%%%%%%%%%%%%%%%%%%%%%%%%%%%%%%%%%%%%%%%%%%%%%%%%%%%%%%%%%%%%%%%%%%%%%%%%%%%%%%%%%%%%%%%%%%%%%%%%%%%%%%%%%%%%%%%%%%%%%%%%%%%%%%%%%%%%%%%%%%%%%%%%%%%%%%%%%%%%%%%%%%%%%%%%%%%%%%%%%%%%%%%%%%%%%%%%%%%%%%%%%%%%%%%%%%%%%%%%%%%%%%%%%%%%%%%%%%%%%%%%%%%%%%%%%%%%%%%%%%%%%%%%%%%%%%%%%%%%%%%%%%%%%%%%%%%%%%%%%%%%%%%%%%%%%%%%%%%%%%%%%%%%%%%%%%%%%%%%%%%%%%%%%%%%%%%%%%%%%%%%%%%%%%%%%%%%%%%%%%%%%%%%%%%%%%%%%%%%%%%%%%%
\subsection{Freezeout with a Chemical Potential}
\label{subsec:chemical}
\begin{figure}[!!!t]
\begin{center}
\includegraphics[width=0.55\textwidth]{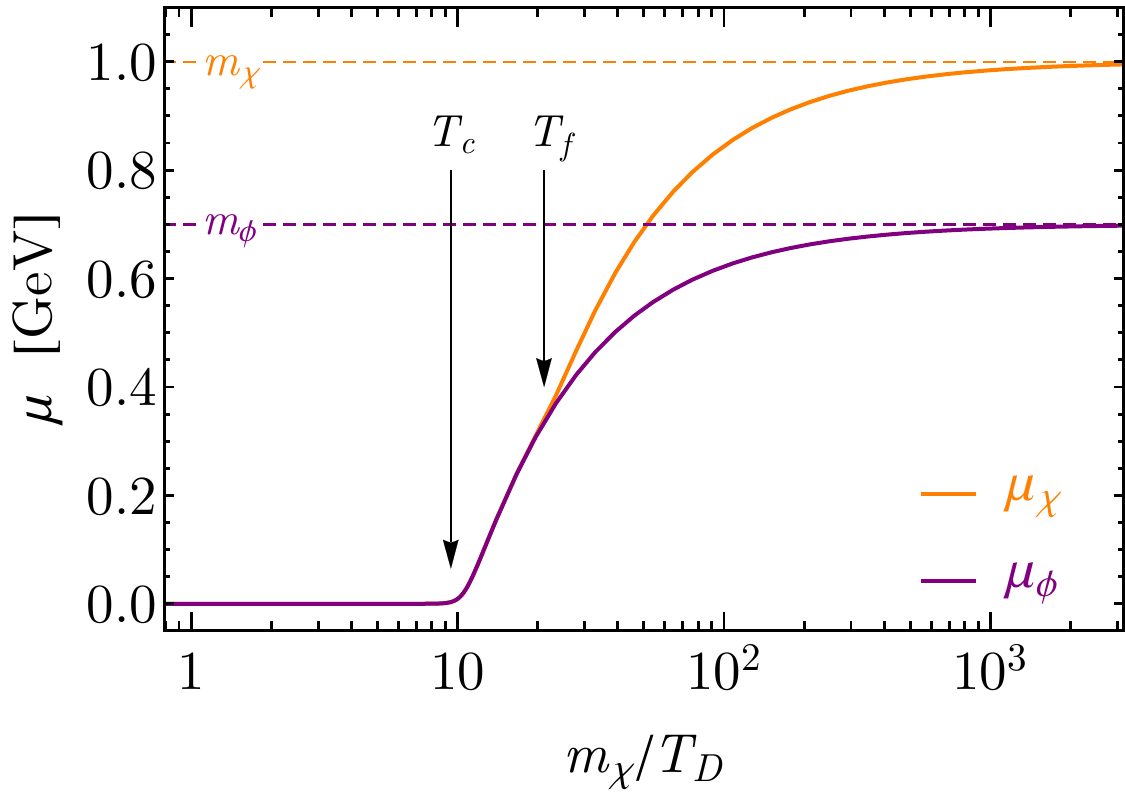}
\end{center}
\vspace{-.3cm}
\caption{\small The temperature evolution of the chemical potentials of $\chi$ and $\phi$.  When $T < T_c$, $\phi \phi \phi \leftrightarrow \phi \phi$  decouples and $\chi$ and $\phi$ develop a chemical potential that encodes conservation of hidden particle number.  Initially, $\mu_\chi = \mu_\phi$ because $\chi \chi \leftrightarrow \phi \phi$ is in equilibrium, but after DM annihilations decouple, $T < T_f$, $\mu_\chi~(\mu_\phi)$ converges to $m_\chi~(m_\phi)$.
}
\label{fig:mu}
\end{figure}

%%%%%%%%%%%%%%%%%%%%%%%%%%%%%%%%%%%%%%%%%%%%%%%%%%%%%%%%%%%%%%%%%%%%%%%%%%
If $T_f< T_c$ (still assuming $T_d$ to be much smaller than both), freeze-out of the $\chi$ number density occurs during the chemical phase, when the total number of dark sector particles is conserved. In particular, if $T_f\ll T_c$, an analytical understanding of the relic density is easily obtained. The relevant Boltzmann equation for this regime is
\be\label{boltzmannchemical}
a^{-3}\frac{d (n_\chi a^3)}{d\log a}=-\frac{\sigma_2}{H}\left(n_\chi^2-\frac{n_\phi^2}{\bar n_\phi^2}\bar n_\chi^2\right).
\ee
Notice that we allow $n_\phi\neq \bar n_\phi$, as number changing interactions have decoupled. Eq.~\ref{boltzmannchemical} is supplemented by the conditions that the total comoving number and entropy densities are conserved for $T<T_c$,
\begin{align}\label{chemNS}
&(n_\phi+n_\chi)a^3\simeq n_\phi a^3\simeq\bar n_\phi a^3\big|_{T_c}\\
&(s_\phi+s_\chi)a^3\simeq s_\phi a^3\simeq \frac{m_\phi-\mu}{T}n_\phi a^3\simeq\bar s_\phi a^3\big|_{T_c},
\end{align}
where the approximate equalities assume $T_c\ll m_\chi-m_\phi$, so that the total number density and entropy are dominated exponentially by $\phi$. As explained in Section~\ref{sec:Thermo}, Eq.~\ref{chemNS} is enforced by the introduction of the chemical potential $\mu$,
\be
n_\phi(T_D)=e^{\mu(T_D)/T_D}\bar n_\phi(T_D),
\ee
which can be calculated approximately using Eq.~\ref{chemNS},
\be\label{mu}
\mu(T_D)=m_\phi\left(1-\frac{T_D}{T_c}\right)~~~{\textrm{and}}~~~ \frac{T_D}{T_c}=\frac{a_c^2}{a^2}.
\ee
The evolution of the chemical potentials of $\chi$ and $\phi$ is shown, for an example parameter point, in Fig.~\ref{fig:mu}.

Given the nonzero chemical potential, Eq.~\ref{boltzmannchemical} implies that if $2\to2$ annihilation are still active in the chemical era, $n_\chi=\bar n_\chi^c=n_\phi \bar n_\chi/\bar n_\phi$ so that
\be\label{nchieqchem}
\bar n^c_\chi(T_D)=2\times \left(\frac{m_\chi T_D}{2\pi}\right)^{3/2} e^{-m_\phi/T_c}e^{-(m_\chi-m_\phi)/T_D},
\ee
where $m_\phi/T_c=x_{0c}$ as in Eq.~\ref{endofcan}. Eq.~\ref{boltzmannchemical} thus reads as the standard Boltzmann equation for a thermal relic in which $\bar n_\chi$ gets replaced by $\bar n^c_\chi$. This implies that the usual condition for sudden freeze-out is replaced by
\be\label{FOchem}
\bar n^c_\chi\, \sigma_2=H.
\ee
During the chemical era, since $T_D\sim 1/a^2$, the Hubble parameter is no longer exponentially sensitive to the dark sector temperature, but
\be\label{HubbleTD}
H(T_D)\simeq  H(T_c)\times 
\begin{cases} 
     T_D/T_c & (\rho_{SM} > \rho_D) \\
(T_D/T_c)^{3/4} & (\rho_{SM} < \rho_D).
   \end{cases}
\ee
The freeze-out temperature obtained from Eq.~\ref{FOchem} reads
\be\label{chemFO}
(1- r) x_f\simeq \log (h(r) m_\chi M_P\sigma_2)-\alpha\log (m_\phi^4 M_P\sigma_3)-\beta \log (x_f)+\gamma\log(x_c)
\ee
where
\be\label{chemh}
h(r)\simeq r^{-1/2-\beta}\times\begin{cases} 
    33\, \xi^{-1/2}g_{*SM}^{1/8}& (\rho_{SM} > \rho_D)\\
   10^3& (\rho_{SM} < \rho_D),
   \end{cases}
\ee
and $\alpha=1/4$, $1/3$, $\beta=1/2$, $3/4$ and $\gamma=0$, $3/4$ depending on whether the SM or the hidden sector dominate the Hubble expansion at freeze-out. The DM yield at freeze-out is given by
\be\label{Ychem}
Y^f_\chi\simeq\frac{1}{\xi} \frac{2\,x^\beta_f x_c^{\gamma-1}}{r^{3/2}h(r)}\times\frac{(m_\phi^4 M_P \sigma_3)^\alpha}{m_\chi M_P \sigma_2}.
\ee
We find that $Y^f_\chi$  is no longer exponentially sensitive to $r$ and it is inversely proportional to $\sigma_2$ as in the case for standard freeze-out. This behavior is shown in Fig.~\ref{Omegar}. Eq.~\ref{Omegacann} for the final relic abundance is almost unchanged in the chemical phase
\be\label{Omegachem}
\frac{\Omega_{\chi}}{\Omega^{\textrm{obs}}_{\rm DM}}\approx 
\begin{cases} 
      0.3\, \frac{x_f}{g_{*SM}^{1/2}} \frac{\sigma_0}{\sigma_2} \frac{T_f}{T_{fSM}}\times \frac{1}{D} & (\rho_{\rm SM} > \rho_D) \\
   0.3\,  \frac{x_f}{g_{*SM}^{1/2}}  \frac{\sigma_0}{\sigma_2} \frac{T_c^{1/2}T_f}{\xi^{1/2} T_{fSM}^{3/2}}\times\frac{1}{D}& (\rho_{\rm SM} < \rho_D)  \\
   \end{cases}
\ee
The only explicit difference with respect to the cannibal phase, Eq.~\ref{Omegacann}, occurs if freeze-out happens during $\phi$ domination, implying that the final abundance in the chemical phase is enhanced by a factor $\sqrt{T_c/T_f}$.

Note that the curves in Fig.~\ref{Omegar} display a change in behavior as $r$ approaches 1. In this region, the approximations of Eq.~\ref{chemNS} are no longer valid. If $m_\chi-m_\phi\ll T_f$, the number density of $\chi$ is no longer exponentially suppressed with respect to that of $\phi$, but $n_\chi\simeq g_{*\chi}/g_{*\phi}\times n_\phi$. In particular this implies that the DM yield is related to the $\phi$ yield at $T_c$,
\be\label{rsim1}
Y_\chi^f\simeq \frac{g_{*\chi}}{g_{*\phi}+g_{*\chi}}\frac{1}{x_c}\frac{1}{\xi},
\ee
which is the result of Ref.~\cite{Carlson:1992fn}. 

%%%%%%%%%%%%%%%%%%%%%%%%%%%%%%%%%%%%%%%%%%%%%%%%%%%%%%%%%%%%%%%%%%%%%%%%%%%%%%%%%%%%%%%%%%%%%%%%%%%%%%%%%%%%%%%%%%%%%%%%%%%%%%%%%%%%%%%%%%%%%%%%%%%%%%%%%%%%%%%%%%%%%%%%%%%%%%%%%%%%%%%%%%%%%%%%%%%%%%%%%%%%%%%%%%%%%%%%%%%%%%%%%%%%%%%%%%%%%%%%%%%%%%%%%%%%%%%%%%%%%%%%%%%%%%%%%%%%%%%%%%%%%%%%%%%%%%%%%%%%%%%%%%%%%%%%%%%%%%%%%%%%%%%%%%%%%%%%%%%%%%%%%%%%%%%%%%%%%%%%%%%%%%%%%%%%%%%%%%%%%%%%%%%%%%%%%%%%%%%%%%%%%%%%%%%%%%%%%%%%%%%%%%%%%%%%%%%%%%%%%%%%%%%%%%%%%%%%%%%%%%%%%%%%%%%%%%%%%%%%%%%%%%%%%%%%%%%%%%%%%%%%%%%%%%%%%%%%%%
\subsection{Freezeout in the Absence of Inverse Annihilations}
\label{subsec:decay}
%%%%%%%%%%%%%%%%%%%%%%%%%%%%%%%%%%%%%%%%%%%%%%%%%%%%%%%%%%%%%%%%%%%%%%%%%%
The discussions of Sections \ref{subsec:cannibal} and \ref{subsec:chemical} assume  the inequality $t_f\ll t_d$ (that is, assumed $\phi$ decays later than DM freeze-out). A careful treatment of what happens dropping this assumption requires solving the full integro-differential Boltzmann equation describing the phase space distribution of $\phi$ and $\chi$. This is beyond the scope of this work.  Here we use a sudden decay approximation, following the strategy outlined in Section~\ref{sec:GenCo}, to estimate the final $\chi$ abundance.  For related studies of the abundance of thermal relics whose annihilation products decay, see Refs.~\cite{Farina:2015uea,Freytsis:2016dgf}.

There is an apparent coincidence of scales in requiring the hidden sector temperature when $\phi$ decays, $T_d$, to fall in between $m_\chi$ and $T_f$, since usually $m_\chi/T_f=O(10)$.  Notice however that if the hidden sector is cannibalizing, an $O(1)$ change in the dark sector temperature corresponds in principle to an exponential variation in Hubble and thus in the timescale. To appreciate this one can calculate the ratio between the age of the Universe at the end of cannibalism and the age of the Universe when $T_D=m_\chi$. This rate scales roughly as,
\be\label{Omegachem}
\frac{t_c}{t(T_D=m_\chi)}\simeq 
\begin{cases} 
10^{-4}\times\xi^{-1/3}(m_\phi^3 M_P\sigma_3)^{1/2} &(\rho_{\rm SM} > \rho_D) \\
10^{-4}\times \xi^{2/3}(m_\phi^3 M_P\sigma_3)^{1/3}& (\rho_{\rm SM} < \rho_D),  \\
   \end{cases}
\ee
depending  on whether cannibalism ends during SM or $\phi$ domination. This ratio can span multiple orders of magnitude, and should be compared with the same quantity evaluated for a hidden sector in equilibrium with radiation. In the latter case, the ratio goes like $(m_\chi/T_f)^2$ and so it is at most $O(10^2)$. 
In the following we will thus only discuss the hierarchy $t_d\ll t_f\ll t_c$. In principle $t_c\ll t_d\ll t_f$ is also possible, but again requires a careful choice for the lifetime of $\phi$. We will thus not discuss this possibility further.

In order to calculate the final DM abundance we follow the sudden decay approximation to the  solution of the Boltzmann system. For $a>a_d$, corresponding to the value of the scale factor when $\phi$ decays, we set $n_\phi=0$ obtaining the following differential equation for $\chi$:
\be\label{boltzmanndecay}
a^{-3}\frac{d (n_\chi a^3)}{d\log a}=-\frac{\sigma_2}{H}\,n_\chi^2.
\ee
It is straightforward to solve this equation approximately, assuming $\sigma_2$ to be constant and $n_\chi(a_d)=\bar n_\chi$. The solution is
\be\label{soldecay1}
n_\chi(a)=\frac{\bar n_\chi(a_d/a)^3}{1+\bar n_\chi\sigma_2/\Gamma_\phi(1-a_d/a)},
\ee
where we used $H(t_d)\sim \Gamma_\phi$.
Notice that since we assume $t_d\ll t_f$, we have $\bar n_\chi \sigma_2\gg \Gamma_\phi$: even though $\phi\phi\to \chi\chi$ processes are not taking place anymore, $\chi\chi\to\phi\phi$ are still active and deplete the $\chi$ abundance. This is completely different from thermal freeze-out, where both forward and inverse processes stop being active at the same time $t_f$, and the residual annihilations occurring for $t>t_f$ only constitute an order one correction to the final abundance. After a short transient Eq.~\ref{soldecay1} becomes
\be\label{soldecay2}
n_\chi(a)=\frac{\Gamma_\phi}{\sigma_2}\left(\frac{a_d}{a}\right)^3.
\ee

The final yield is obtained dividing the number density by the entropy density which, using Eq.~\ref{TRH}, can always be written as
\be
s_{SM}\left(\frac{a}{a_d}\right)^3=\frac{2\pi^2}{45}g_{*SM}T_{RH}^3\approx 0.07\, g_{*SM}^{1/4}(\Gamma_\phi M_P)^{3/2},
\ee
independently of whether $\phi$ is dominating or not the energy density of the Universe when it decays. We have
\be\label{Omegadecay}
\frac{\Omega_\chi}{\Omega^{\textrm{obs}}_{DM}}\simeq\frac{0.9}{g_{*SM}^{1/2}\,r\,\gamma_\phi^{1/2}}\frac{\sigma_0}{\sigma_2},
\ee
where we define $\gamma_\phi\equiv\Gamma_\phi/H\big|_{T_{SM}=m_\phi}$ and we approximate $g_{*SM}$ as constant. By Eq.~\ref{outofeq}, $\gamma_\phi\ll1$ in order for the hidden sector and the SM not to be in thermal equilibrium. This generally implies a boosted $2\to2$ cross section to reproduce the right relic abundance. Notice that the boost increases as $\Gamma_\phi$ gets smaller. This behavior holds until $t_d\approx t_f$, at which point the boost saturates and $\Gamma_\phi$ enters only through the dilution factor if $\phi$ dominates the energy density of the Universe.

\begin{table}
\begin{center}
{\renewcommand{\arraystretch}{1.75} 
\begin{tabular}{|c|c|c|}  
\hline
 phase & $\rho_D < \rho_{SM}$ & $\rho_D > \rho_{SM}$ \\ \hline \hline
cannibal &   {\footnotesize $D^{-1}$}$\frac{T_f}{T_{fSM}} $  & {\footnotesize $D^{-1}$}$ \frac{T_f^{3/2}}{\xi^{1/2}T_{fSM}^{3/2}}$ \\ \hline
chemical &  {\footnotesize $D^{-1}$}$\frac{T_f}{T_{fSM}} $ &  {\footnotesize $D^{-1}$}$ \frac{T_c^{1/2}T_f}{\xi^{1/2}T_{fSM}^{3/2}}$ \\ \hline
one way & \multicolumn{2}{c|}{ $\gamma_\phi^{-1/2}$ } \\ \hline 
\end{tabular}
}
\end{center}
  \caption{ \label{tab:boost} \small
The boost in the DM annihilation rate for each phase (cannibal, chemical, or one way), depending on whether the energy density is dominated by the SM ($\rho_D < \rho_{SM}$) or the dark sector ($\rho_D > \rho_{SM}$) when DM annihilations decouple at time $t_f$.  The DM annihilation rate is boosted above the thermal WIMP value ($\sigma_0 = 3 \times 10^{-26}~\mathrm{cm}^2/\mathrm{s}$) by a multiplicative factor proportional to the value shown in the table.
  }
\end{table}

To summarize this section, the parametric form of the DM relic density is modified in each phase (see Eqs.~\ref{Omegacann}, \ref{Omegachem}, and \ref{Omegadecay}), compared to the conventional WIMP case of Eq.~\ref{eq:VanillaOmega}.  In order to match the observed relic density, the DM annihilation rate is generically boosted above the conventional s-wave thermal WIMP value of $\sigma_0 = 3 \times 10^{-26}~\mathrm{cm}^3/\mathrm{s}$.  The parametrics of the boost factor for each phase are collected in Tab.~\ref{tab:boost}.

%%%%%%%%%%%%%%%%%%%%%%%%%%%%%%%%%%%%%%%%%%%%%%%%%%%%%%%%%%%%%%%%%%%%%%%%%%%%%%%%%%%%%%%%%%%%%%%%%%%%%%%%%%%%%%%%%%%%%%%%%%%%%%%%%%%%%%%%%%%%%%%%%%%%%%%%%%%%%%%%%%%%%%%%%%%%%%%%%%%%%%%%%%%%%%%%%%%%%%%%%%%%%%%%%%%%%%%%%%%%%%%%%%%%%%%%%%%%%%%%%%%%%%%%%%%%%%%%%%%%%%%%%%%%%%%%%%%%%%%%%%%%%%%%%%%%%%%%%%%%%%%%%%%%%%%%%%%%%%%%%%%%%%%%%%%%%%%%%%%%%%%%%%%%%%%%%%%%%%%%%%%%%%%%%%%%%%%%%%%%%%%%%%%%%%%%%%%%%%%%%%%%%%%%%%%%%%%%%%%%%%%%%%%%%%%%%%%%%%%%%%%%

\section{Probing Cannibal Dark Matter}
\label{sec:pheno}
In this section, we study the observational constraints and future reach to probe cannibal DM. As a consequence of the out of equilibrium condition, Eq.~\ref{outofeq}, cannibal DM is unobservable in direct detection experiments. Astrophysical and cosmological signatures are however possible and they stem from three characteristic features of our framework:
\begin{itemize}
\item a boosted dark matter annihilation rate (see Tab.~\ref{tab:boost}),
\item the existence of a long lived LDP,
\item a large temperature ratio between the dark sector and the SM at DM freeze-out.
\end{itemize}
Even though our discussion will take place within the model introduced in Section~\ref{sec:GenCo}, these properties are completely general and our conclusions can easily be generalized to other implementations.

\begin{figure}[!!!t]\hbox{\hspace{-0.8cm}
\includegraphics[width=0.49\linewidth]{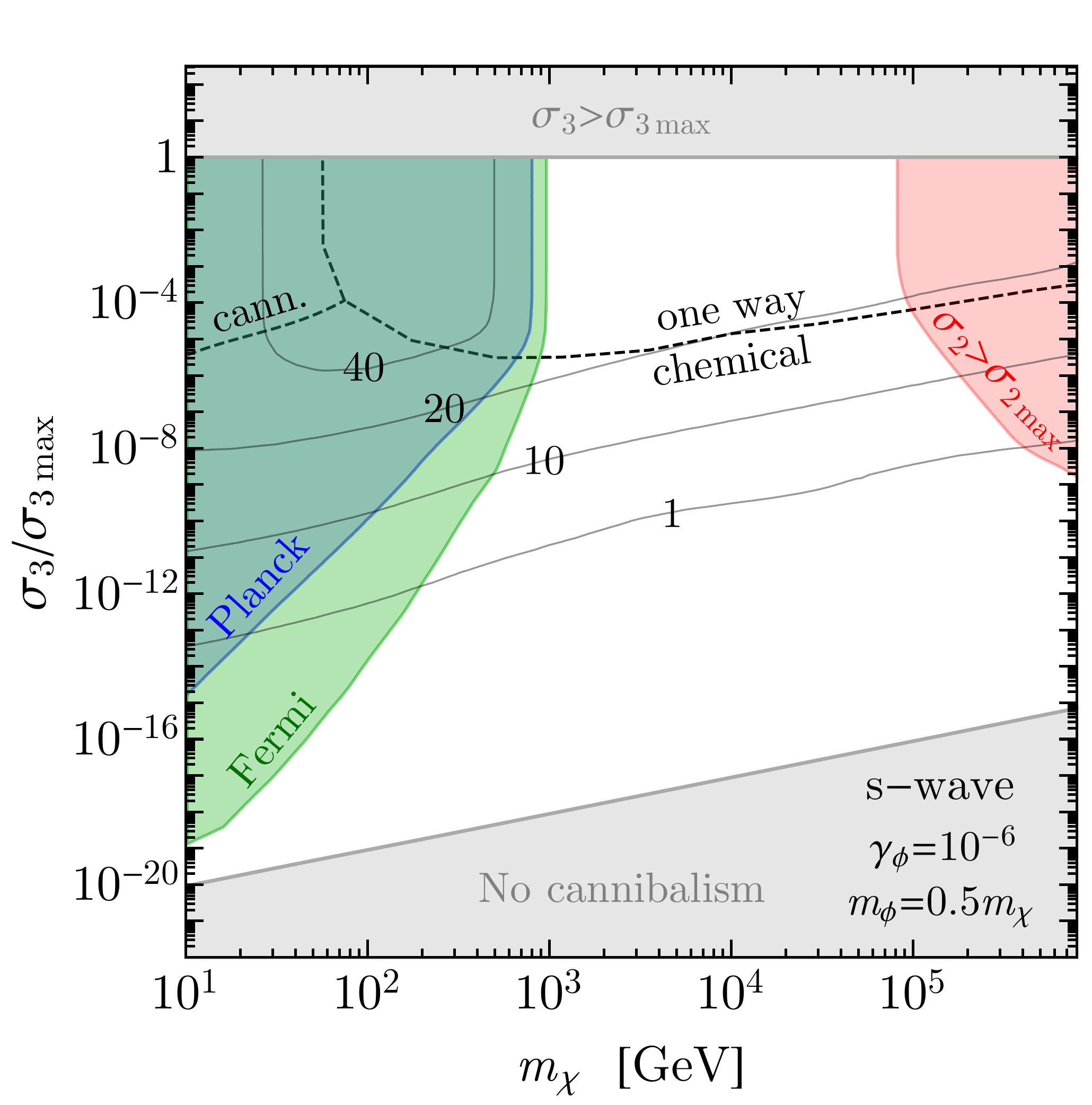}~~~~~\includegraphics[width=0.49\linewidth]{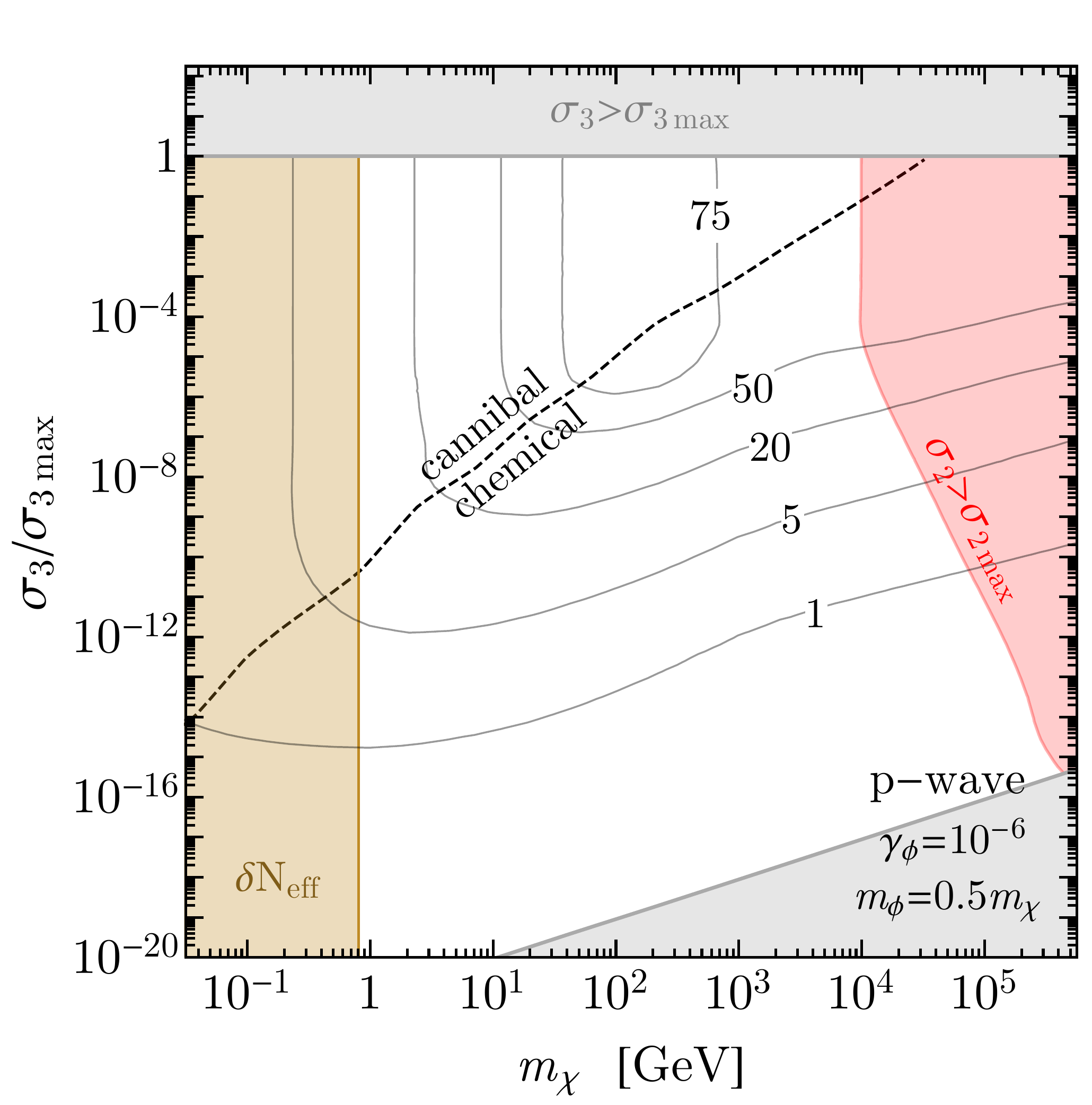}}
\caption{\label{figpheno1} \small
Observational reach and constraints on cannibal DM for s-wave ({\it left}) and p-wave ({\it right}) annihilations, as a function of the DM mass, $m_\chi$, and the $3\phi \rightarrow 2 \phi$ cross section, $\sigma_3$, normalized to the perturbativity limit $\sigma_{3\rm max}$.  We set $m_\phi = 0.5 m_\chi$ and the $\phi$ lifetime is chosen such that $\gamma_\phi \equiv \Gamma_\phi / H(T_{SM} = m_\phi) = 10^{-6}$.   At each point in the plane we choose the $\chi \chi \rightarrow \phi \phi$ annihilation rate, $\sigma_2$, such that $\Omega_\chi$ matches the observed DM relic density.    The gray contours describe the multiplicative boost factors $B_s$ and $B_p$ to the DM annihilation rate, relative to the usual thermal WIMP value, as defined in the text.  The dashed black line delineates the transition between the cannibal, chemical, and one way phases.  In the upper gray region, $\sigma_3$ becomes non-perturbative, and in the lower gray region, $3 \phi \rightarrow 2 \phi$ decouples when $T_D > m_\phi$, such that the hidden sector never undergoes cannibalism.  In the red shaded region, $\sigma_2$ becomes non-perturbative.  The shaded blue (green) region is excluded by Fermi (Planck) constraints on $2 \chi \rightarrow 2 \phi \rightarrow 4 \gamma$~\cite{Ackermann:2015zua,Ade:2015xua,Slatyer:2015jla,Elor:2015bho}.  The shaded brown region is excluded by the Planck constraint on $N_{\textrm{eff}}$~\cite{Ade:2015xua}, because $\phi \rightarrow 2 \gamma$ heats photons relative to neutrinos, lowering $N_{\textrm{eff}}$.
}
\end{figure}

\begin{figure}[!!!t]\hbox{\hspace{-0.8cm}
\includegraphics[width=0.49\linewidth]{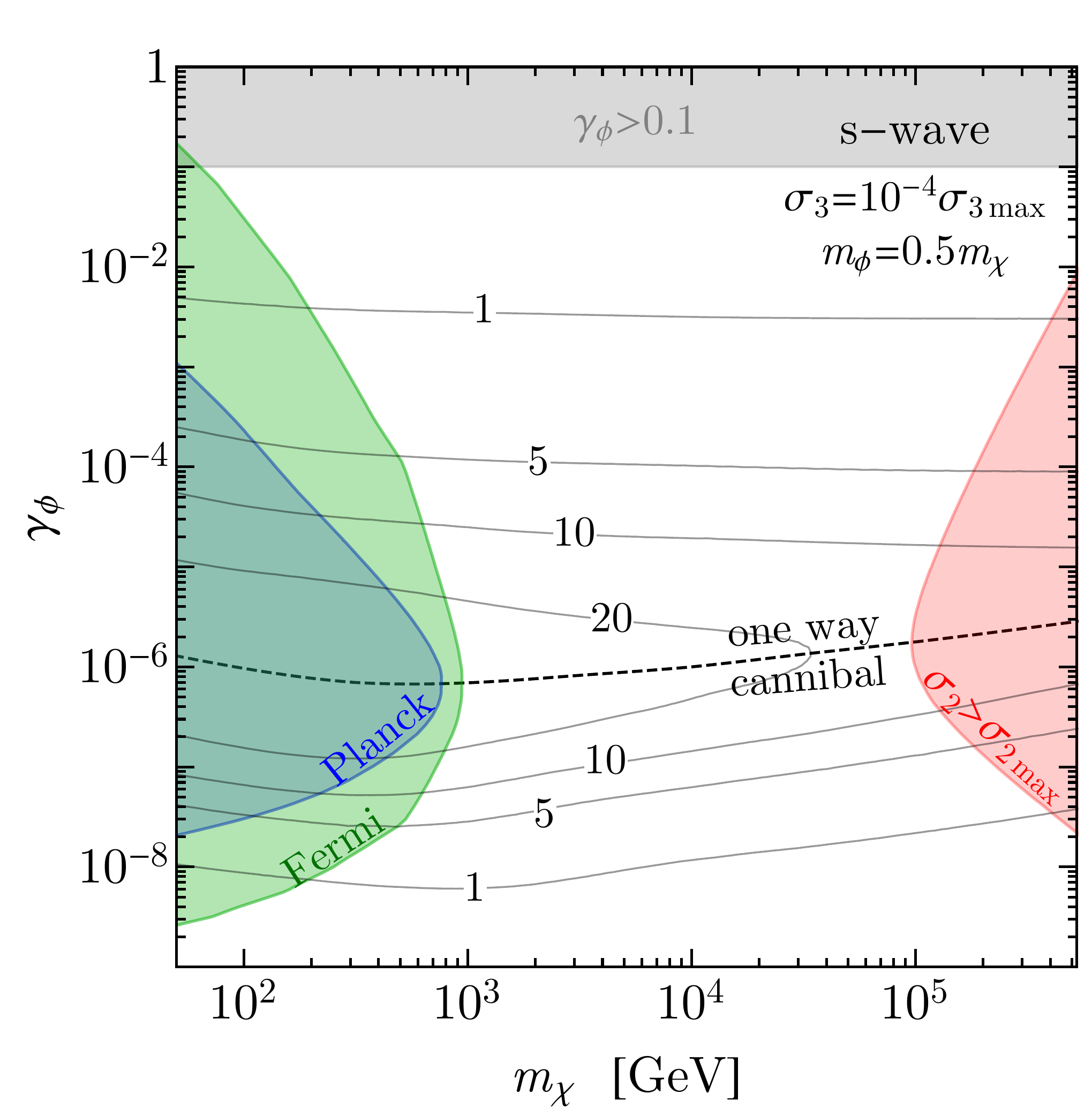}~~~~~\includegraphics[width=0.49\linewidth]{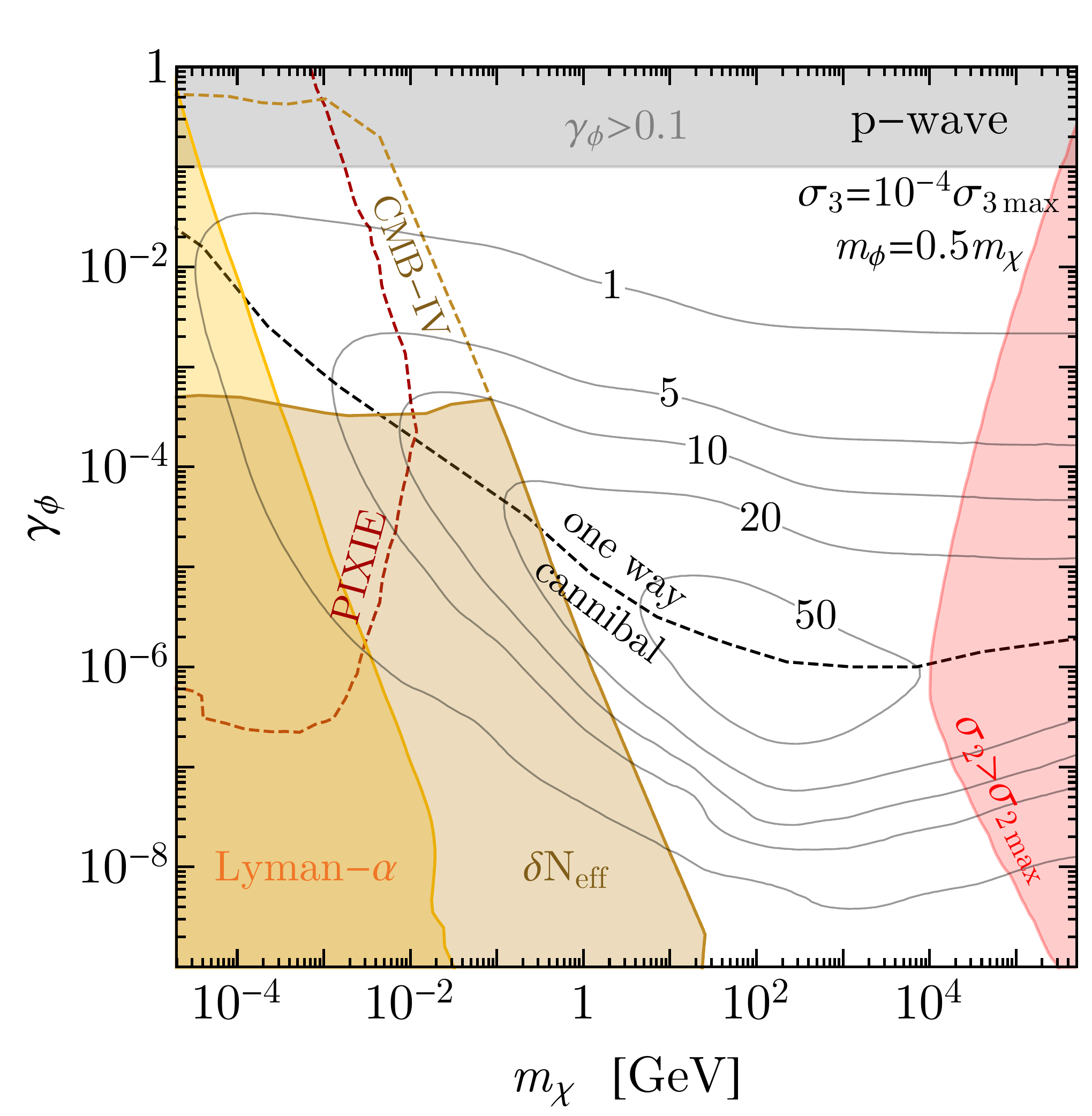}}
\caption{\label{figpheno2} \small
Observational reach and constraints on cannibal DM for s-wave ({\it left}) and p-wave ({\it right}) annihilations, as a function of the DM mass, $m_\chi$ and the $\phi$ lifetime, $\gamma_\phi \equiv \Gamma_\phi / H(T_{SM} = m_\phi)$.  We set $m_\phi = 0.5 m_\chi$ and the $\phi \phi \phi \rightarrow \phi \phi$ cross section to $\sigma_3 = 10^{-4} \sigma_{3\rm max}$, where $\sigma_{3\rm max}$ is the perturbative limit.  At each point in the plane we choose the $\chi \chi \rightarrow \phi \phi$ annihilation rate, $\sigma_2$, such that $\Omega_\chi$ matches the observed DM relic density.  The gray contours describe the multiplicative boost factors $B_s$ and $B_p$ to the DM annihilation rate, relative to the usual thermal WIMP value, as defined in the text.
The yellow region is excluded by Lyman-$\alpha$ constraints on the DM free-streaming length~\cite{Viel:2013apy,Baur:2015jsy}.  The dashed maroon curve shows the reach of PIXIE~\cite{PIXIE} to measure spectral distortions to the CMB caused by $2\chi \rightarrow 2 \phi \rightarrow 4 \gamma$.  The dashed brown curve shows the reach of planned CMB Stage-IV experiments~\cite{Wu:2014hta} to measure a deviation to $N_{\textrm{eff}}$ because $\phi \rightarrow 2 \gamma$ heats photons relative to neutrinos.
The remaining constraints (shaded) and reach (dashed) curves are the same as in Fig.~\ref{figpheno1}.
}
\end{figure}

In order to investigate the observational constraints and reach, we study the $(m_\chi, \sigma_3)$ and $(m_\chi, \gamma_\phi)$ planes in Figs.~\ref{figpheno1} and \ref{figpheno2}, respectively.   For each figure we study $s$-wave ($p$-wave) annihilations for DM on the left (right).   At each point in these planes, we choose the $\chi \chi \rightarrow \phi \phi$ cross section, $\sigma_2$, such that the DM relic density matches observation.  We find that the DM annihilation rate is generically boosted above typical value for a thermal WIMP\@.  In order to describe the cross section, we define multiplicative boost factors as follows,
\begin{align}
&\left< \sigma v \right> \approx B_s \times \left( 3 \times 10^{-26}~\mathrm{cm}^3 / \mathrm{s} \right)  & (s-\rm{wave}) \\
&\left< \sigma v \right> \approx B_p \times \left( 1 \times 10^{-24}~\mathrm{cm}^3 / \mathrm{s} \right) \left( \frac{m_\chi}{T_D} \right)  &(p-\rm{wave})
\end{align}
Figs.~\ref{figpheno1} and \ref{figpheno2} show contours of $B_s$ and $B_p$, which vary from about 1 to 75.  Here we have fixed  $\xi = \xi_0 \approx 39$, but note that varying $\xi$ can allow for significantly larger boost factors.

{\bf{Indirect detection and CMB energy injection}} - DM annihilations at late times are constrained by both Fermi, through the observation of gamma rays coming from Dwarf Spheroidal galaxies surrounding the Milky way~\cite{Ackermann:2015zua}, and by CMB measurements performed by Planck, that can constrain late time energy injection in the SM plasma affecting the ionization history of the Universe~\cite{Ade:2015xua}. In our case $\chi\chi\to\phi(\to\gamma\gamma)\phi(\to\gamma\gamma)$.
For our $s$-wave model we use the bounds presented in~\cite{Ade:2015xua,Slatyer:2015jla,Elor:2015bho}. Constraints on late time annihilation are typically much less stringent for $p$-wave annihilations. 

{\bf{Number of relativistic degrees of freedom}} - If $\phi$ decays to SM radiation after the time of neutrino decoupling, its energy density will increase the temperature of photons with respect to that of neutrinos, lowering $N_{\textrm{eff}}$, the effective number of relativistic degrees of freedom.\footnote{In~\cite{cannibal} the case in which $\phi$ decays to decoupled radiation, e.g. a dark photon, was studied. In that case $N_{\textrm{eff}}$ is increased by the energy released when $\phi$ decays.} $N_{\textrm{eff}}$ is constrained by CMB measurements performed by Planck~\cite{Ade:2015xua} which bound $N_{\textrm{eff}}=3.15\pm0.23$. 
We also include the projected sensitivity on $N_{\textrm{eff}}$ from CMB Stage-IV experiments for which $\delta N_{\textrm{eff}}=0.03$ at 95\% confidence level~\cite{Wu:2014hta}.  We find that the present $N_{\textrm{eff}}$ bound excludes the possibility of $\phi$ dominating the energy density at the beginning of BBN, so that the prediction of standard BBN for the abundance of helium and deuterium are not modified (through possible nonstandard evolution of the Hubble parameter) once the $N_{\textrm{eff}}$ bound is satisfied. In the allowed region in the right panel of Fig.~\ref{figpheno1}, $\phi$ decays before BBN so there is no reach from CMB-IV experiments. We also find that once the $N_{\textrm{eff}}$ bound is satisfied, energy injection in the plasma by $\phi\to\gamma\gamma$ decays does not affect the abundance of light elements as predicted by standard BBN \cite{Jedamzik:2006xz}.

{\bf{CMB spectral distortion}} - Late annihilations $\chi \chi \rightarrow \phi \phi$ followed by $\phi\to\gamma\gamma$ decay are constrained by the shape of the phase space distribution of the CMB photons~\cite{McDonald:2000bk}. When the SM temperature drops below $T_{DC}\approx1\,\textrm{keV}$, double-Compton scatterings become inefficient and the total comoving number of photons is approximately conserved. Since Compton scatterings are still active, energy exchange between the photons is still possible (at least down to $T_{C}\approx 1\,{\textrm{eV}}$). 
The photons injected in the SM plasma by the residual $\chi \chi \rightarrow \phi(\to\gamma\gamma) \phi(\to\gamma\gamma)$ annihilations occurring for $T_{C}<T_{SM}<T_{DC}$ give rise to a non-vanishing chemical potential for the photon, generating a so-called $\mu$-distortion of the CMB spectrum~\cite{Chluba:2011hw}. The best bound on this effect was set by FIRAS to be $\mu<9\times10^{-5}$ at 95\% confidence level~\cite{Fixsen:1996nj}. While the current bound is never relevant to us, we include the reach of future experiments like PIXIE, which are sensitive to sub-GeV masses, that projects sensitivity to $\mu \gtrsim 10^{-8}$~\cite{PIXIE}.

{\bf{Free-streaming}} - As DM kinetically decouples from the thermal bath, it starts to freely diffuse across the Universe. This effect suppresses matter perturbations below the so-called free-streaming scale, defined as the comoving distance traveled by a DM particle between the time of kinetic decoupling and the time of matter-radiation equality.
In our model, thermal decoupling for $\chi$ happens when the rate of elastic scattering, $\chi\phi\to\chi\phi$, falls below the Hubble rate
\be\label{kineticdec}
n_\phi \langle\sigma_{\textrm{el}}\rangle < H.
\ee
Defining $\bar T_{\textrm{k}}$ as the dark sector temperature at which Eq.~\ref{kineticdec} is realized in the limit $\Gamma_\phi=0$, we calculate the kinetic decoupling temperature $T_{\textrm{k}}$ in our setup as $T_{\textrm{k}}=\max( T_d, \bar T_{\textrm{k}})$. In most of the parameter space of our model, kinetic decoupling happens when $\phi$ decays, setting $n_\phi=0$ in our sudden decay approximation. In this case, we can give a simple formula for the free-streaming length:
\be\label{freestreaming}
\lambda_{\textrm{fs}}=\int_{t_{\rm k}}^{t_{\rm eq}}\frac{v(t)}{a(t)}dt\approx 125\,{\textrm{Mpc}}\, v_{\textrm{k}}\frac{\log (1.3\,T_{\textrm k}^{SM}/\text{eV})}{T_{\textrm k}^{SM}/{\textrm{eV}}},
\ee
where $v_k\sim\sqrt{T_{\textrm{k}}^D/m_\chi}$ and $T_{\textrm{k}}^{SM}$ are the velocity of dark matter and the temperature of the SM, respectively, at the kinetic decoupling of the DM from its thermal bath. This free-streaming length will generally be larger than for a regular WIMP, due to the fact that the hidden sector underwent an era of cannibalism resulting in $T_{\textrm{k}}^D>T_{\textrm{k}}^{SM}$ and $v_{\textrm{k}}>v_{\textrm{k}}^{WIMP}$. The strongest constraint on $\lambda_{\textrm{fs}}$ comes from measurements of the Lyman-$\alpha$ forest spectra, implying $\lambda_{\text{fs}}\lesssim0.1 \text{ Mpc}$~\cite{Viel:2013apy,Baur:2015jsy}.

Each of these constraints and future reach are displayed in Figs.~\ref{figpheno1} and \ref{figpheno2}.  We find a sizable allowed parameter space, for each of the three phases (cannibal, chemical, and one way), where the DM annihilation rate is boosted above the prediction for a typical thermal WIMP\@.  The right panel of Fig.~\ref{figpheno2} shows that significant parameter space can be discovered by CMB Stage-IV measurements of $N_{\textrm{eff}}$ and PIXIE measurements of $\mu$-distortions.  We note that it would be interesting for future studies of cannibal DM to explore more initial conditions for $\xi$, to consider more SM final states beyond the $2 \chi \rightarrow 2 \phi \rightarrow 4 \gamma$ case considered here, and to consider the case that $\phi$ decays into dark radiation instead of the SM (as was considered for the cannibal phase in Ref.~\cite{cannibal}).

%%%%%%%%%%%%%%%%%%%%%%%%%%%%%%%%%%%%%%%%%%%%%%%%%%%%%%%%%%%%%%%%%%%%%%%%%%

{\it Note added:}  While completing this work, we became aware of Ref.~\cite{Cornell}, which considers the one way phase for the non-generic spectrum $m_\phi \approx m_\chi$.  They consider both the limit $t_c \rightarrow t_\phi$ and the ordering $t_c > t_d$.

\section*{Acknowledgments}
We thank Jens Chluba, Raffaele D'Agnolo, Jeff Asaf Dror, and Maxim Pospelov for helpful discussions.  M.F. is supported in part by the DOE Grant DE-SC0003883 and D.P. and G.T. are supported by the James Arthur Postdoctoral Fellowship.  This work was supported in part by the hospitality of the Aspen Center for Physics, which is supported by National Science Foundation grant PHY-1066293. 
%%%%%%%%%%%%%%%%%%%%%%%%%%%%%%%%%%%%%%%%%%%%%%%%%%%%%%%%%%%%%%%%%%%%%%%%%%%%%%%%%%%%%%%%%%%%%%%%%%%%%%%%%%%%%%%%%%%%%%%%%%%%%%%%%%%%%%%%%%%%%%%%%%%%%%%%%%%%%%%%%%%%%%%%%%%%%%%%%%%%%%%%%%%%%%%%%%%%%%%%%%%%%%%%%%%%%%%%%%%%%%%%%%%%%%%%%%%%%%%%%%%%%%%%%%%%%%%%%%%%%%%%%%%%%%%%%%%%%%%%%%%%%%%%%%%%%%%%
\appendix
%%%%%%%%%%%%%%%%%%%%%%%%%%%%%%%%%%%%%%%%%%%%%%%%%%%%%%%
\section{Boltzmann equations}\label{boltzmann}
As explained in the text, our sudden decay approximation requires us to solve two different systems of Boltzmann equations depending on whether $t<t_d$ or $t>t_d$, where $t_d$ is defined as $t_d\equiv H^{-1}=\Gamma_\phi^{-1}$.
For $t<t_d$ we have
\be\label{system}
a^3 H \frac{d\,n_\phi a^3}{d \log a}=\mathcal K_{\phi},~~~~~a^3 H \frac{d\,n_\chi a^3}{d \log a}=\mathcal K_{\chi},
\ee
and an equation for the entropy conservation in the dark sector
\be\label{darkentropy}
a^3 H \frac{d(s_\phi+s_\chi) a^3}{d \log a}=0.
\ee
Starting from the phase space distribution of a particle species in thermal equilibrium in the hidden sector,
\be\label{phasedist}
f_X(p,T_D)=e^{\mu_X(T)/T_D}e^{E(p)/T_D}=\frac{n_X(T)}{\bar n_X(T)}e^{E(p)/T_D},
\ee
the associated entropy density is given by
\be\label{entropydef}
s_X=\int\frac{d^3p}{(2\pi)^3}\, f_X(p)(1-\log f_X(p))=\frac{\rho_X-\mu_X n_X}{T_D}+n_X,
\ee
where $\rho_X$ and $n_X$ have their usual definitions. Using Eq.~\ref{entropydef}, the two entropy conservation equations in the first line of Eq.~\ref{darkentropy} can be turned into an equation for the evolution of the dark sector temperature as a function of the scale factor. 

The two kernels in Eq.~\ref{system} are obtained by integrating the appropriate Boltzmann equation
\begin{align}\label{kernels}
\mathcal K_X=\sum \int d\Pi_Xd\Pi_a\ldots d\Pi_{\bar a}\ldots\times& (2\pi)^4\delta^{(4)}(p_X+p_a+\ldots-p_{\bar a}+\ldots)\\\nonumber
&\times S_X |\mathcal M_{Xa\ldots\to\bar a\ldots}|^2\times [f_Xf_a\ldots-f_{\bar a}\ldots],
\end{align}
where $d\Pi=d^3 p/[2E(2\pi)^3]$ and the sum is over all the reactions involving $X$. The factor $S_X$ is a symmetry factor that counts the number of identical $X$ particles in the initial state. The various squared matrix element in Eq.~\ref{kernels} are averaged over initial and final state quantum numbers; furthermore they include the appropriate symmetry factor for all identical initial and final state particles. The kernels $\mathcal K_X$ can be rewritten as
\be
\mathcal K_X=\sum\Delta_X \gamma_{Xa\ldots\to \bar a\ldots}\left(\frac{n_X}{\bar n_X}\frac{n_a}{\bar n_a}\ldots -\frac{n_{\bar a}}{\bar n_{\bar a}}\ldots\right),
\ee
where $\gamma_{Xa\ldots\to \bar a\ldots}$ is the rate per unit volume of the $Xa\ldots\to \bar a\ldots$ reaction
\begin{align}
\gamma_{Xa\ldots\to \bar a\ldots}=S_X\int (d\Pi_X \bar f_X)(d\Pi_a\bar f_a)\ldots \int d\Pi_{\bar a}\ldots(2\pi)^4\delta^{(4)}(p_X+\ldots-p_{\bar a}+\ldots)|\mathcal M_{Xa\ldots\to\bar a\ldots}|^2,
\end{align}
and $\Delta_X$ counts by how many units the process $Xa\ldots\to \bar a\ldots$ changes the number of $X$ particles. The reaction rates are typically written in terms of thermally averaged cross sections as
\be
\gamma_{Xa\ldots\to \bar a\ldots}\equiv \langle\sigma_{Xa\ldots\to \bar a\ldots} v^n \rangle \bar n_X\bar n_a\ldots
\ee
where $n=1$ for $2\to 2$ reactions and $n=2$ for $3\to 2$ ones. 
The only $2\to 2$ process we include in our calculation is $\chi\chi\rightarrow\phi\phi$. 
Among the various $3\to2$ ones, the only relevant ones are $\phi\phi\phi\to \phi\phi$ and $\phi\phi\phi\to \chi\chi$.

Neglecting subleading corrections of order $T/m_\chi$,
\be\label{chichiswave}
\langle\sigma_{\chi\chi\to\phi\phi} v\rangle\approx \frac{y_5^2\sqrt{1-r^2}}{128\pi m_\chi^2}\left[\frac{4y(4-r^2)+a_\phi r(2-r^2)}{8-6 r+r^4}\right]^2
\ee
for $s$-wave and 
\begin{equation}\label{chichipwave}
\langle\sigma_{\chi\chi\to\phi\phi} v\rangle\equiv\frac{T}{m_\chi} \sigma_0 \approx\frac{T}{m_\chi}\frac{\sqrt{1-r^2}}{128\pi m_\chi^2}\left[\frac{3 a_\phi^2 r^2}{(4-r^2)^2}+\frac{8a_\phi y\, r(20-13r^2+2r^4)}{(4-r^2)(2-r^2)^2}+\frac{16 y^2(9-8r^2+2r^4)}{(2-r^2)^4}\right]
\end{equation}
for $p$-wave. We defined $a_\phi=A/m_\phi$. Notice that for $r=1$, the cross sections vanish spuriously. In this limit, the corrections of order $T/m_\chi$ to the thermal average cannot be neglected. Keeping these subleading corrections, the averaged cross sections become suppressed by a factor $\sqrt{T/m_\chi}$ in the $1-r\ll T/m_\chi$ limit. We keep the full $T$ dependence in our codes.

For the $3\to2$ processes, we have
\be\label{3phi2phi}
\langle\sigma_{\phi\phi\phi\to\phi\phi} v^2\rangle\approx \frac{25\sqrt 5 a_\phi^2(a_\phi^2+3\lambda)^2}{331776\pi m_\phi^5},
\ee
and
\be\label{3phi2chi}
\langle\sigma_{\phi\phi\phi\to\chi\chi} v^2\rangle\approx \frac{9(y^2+y_5^2)r^2-4 y^2}{221184\pi  r^{11}m_\chi^{5}}\left[48 y^2-16 a_\phi y r-(\lambda+12(y^2+y_5^2)-a^2)r^2\right]^2\sqrt{1-\tfrac{4}{9r^2}}.
\ee
Eq.~\ref{3phi2chi} only holds if $3m_\phi-2m_\chi\gg T$. We use the full thermal average in our code. 
%%%%%%%%%%%%%%%%%%%%%%%%%%%%%%%%%%%%%%%%%%%%%%%%%%%%%%%%%%%%%%%%%%%%%%%%%%

%%%%%%%%%%%%%%%%%%%%%%%%%%%%%%%%%%%%%%%%%%%%%%%%%%%%%%%%%%%%%%%%%%%%%%%%%%%%%%%%%%%%%%%%%%%%%%%%%%%%%%%%%%%%%%%%%%%%%%%%%%%%%%%%%%%%%%%%%%%%%%%%%%%%%%%%%%%%%%%%%%%%%%%%%%%%%%%%%%%%%%%%%%%%%%%%%%%%%%%%%%%%%%%%%%%%%%%%%%%%
%%%%%%%%%%%%%%%%%%%%%%%%%%%%%%%%%%%%%%%%%%%%%%%%%%%%%%%%%%%%%%%%%%%%%%%%%%%%%%%%%%%%%%%%%%%%%%%%%%%%%%%%%%%%%%%%%%%%%%%%%%%%%%%%%%%%%%%%%%%%%%%%%%%%%%%%%%%%%%%%%%%%%%%%%%%%%%%%%%%%%%%%%%%%%%%%%%%%%%%%%%%%%%%%%%%%%%%%%%%%%%%%%%%%%%%%%%%%%%%%%%%%%%%%%%%%%%%%%%%%%%%%%%%%%%%%%%%%%%%%%%%%%%%%%%%%%%%%%%%%%%%%%%%%%%%%%%%%%%%%%%%%%%%%%%%%%%%%%%%%%%%%%%%%%%%%%%%%%%%%%%%%%%%%%%%%%%%%%%%%%%%%%%%%%%%%%%%%%%%%%%%%%%%%%%%%%%%%%%%%%%%%%%%%%%%%%%%%%%%%%%%%%%%%%%%%%%%%%%%%%%%%%%%%%%%%%%%%%%%%%%%%%%%%%%%%%%%%%%%%%%%%%%%%%%%%%%%%%%%%%%%%%%%%%%%%%%%%%%%%%%%%%%%%%%%%%%%%%%%%%%%%%%%%%%%%%%%%%%%%%%%%%%%%%%%%

\end{document}